\documentclass{aastex}         
\usepackage{spr-astr-addons}    
\usepackage{url}
\usepackage{hyperref}
\usepackage{xcolor}
\hypersetup{
    colorlinks,
    linkcolor={red!50!black},
    citecolor={blue!50!black},
    urlcolor={blue!80!black},
}
\usepackage{graphicx}
\batchmode
\renewcommand{\arraystretch}{1.25}
\renewcommand{\tabcolsep}{1.250mm}

\usepackage{amsmath}
\usepackage{amsfonts}
\usepackage{amssymb}
\usepackage{breqn}

\begin{document}
%
\title{Halo Orbits around $L_1$ and $L_2$ in the Photogravitational Sun-Earth System with Oblateness}

\shorttitle{Halo Orbits around $L_1$ and $L_2$}
\shortauthors{Dhwani, Thomas}

\author{Dhwani Sheth\altaffilmark{1} }
\email{dhwani.sheth-mathphd@msubaroda.ac.in}

\and
\author{V. O. Thomas\altaffilmark{1} }
\email{votmsu@gmail.com} 
\altaffiltext{1}{Department of Mathematics, Faculty of Science, The Maharaja Sayajirao University of Baroda, Vadodara-390002, India\\email: dhwani.sheth-mathphd@msubaroda.ac.in}

\begin{abstract}
The Photogravitational Restricted Three Body Problem with oblateness has been studied to obtain halo orbits around the Lagrangian points $L_1$ and $L_2$ of the Sun-Earth system in which the Sun is taken as radiating and the Earth as an oblate spheroid. 
The halo orbits corresponding to fourth and fifth order approximations around $L_1$ and $L_2$ for actual oblateness of the Earth and for different radiation pressures for the Sun are displayed graphically. 
The time period of halo orbits around $L_1$  decreases with increase in oblateness and increases with increase in radiation pressure. A reverse effect is observed due to increase in oblateness and radiation pressure on time period of orbits around $L_2$. It is also observed that 
halo orbits around $L_1$ shifts towards the source of radiation due to increase in both radiation pressure and oblateness. However, halo orbits around $L_2$ shifts towards the source of radiation due to increase in radiation but recedes with increase in oblateness. 
\end{abstract}


\section{Introduction}
	Restricted Three Body Problem(RTBP) deals with the motion of an infinitesimal body which moves under the gravitational influence of two massive bodies called the primaries. The infinitesimal body is called the secondary body. The only force acting on this system is the gravitational attraction force between the primaries. The mass of the secondary body is negligible compared to the primary masses and it does not influence the motion of the primaries.
	RTBP is very useful for describing the motion of planets, asteroids, comets and satellites\citep{Plummer,Winter,Brouwer&Clemence,Danby,Pollard,Murray&Dermot}. It plays an important role in space dynamics, celestial dynamics and analytic dynamics. It has applications in the fields of mathematics, theoretical physics and quantum physics. In Circular Restricted Three Body Problem(CRTBP), the primaries move in a circular path around their common centre of mass. This is a particular case of RTBP\citep{Moulton,McCuskey,Szebehely,Roy,Fitzpatrick,Vallado}.
	Most of the celestial bodies are radiating and hence study of RTBP incorporating radiation, usually called photogravitational RTBP, is pertinent. 
	The solar radiation pressure force changes with the distance in a similar law as the gravitational attraction force but acts in an opposite direction to it. This reduces the effective mass of the Sun\citep{Poynting1903,Robertson,Schuerman,Simmons,Abouelmagd2013,Niraj}.
	In the case of planar CRTBP, there exist five equilibrium points known as Libration points or Lagrangian points. Among these, three points, denoted by $L_1, L_2$ and $L_3$ are collinear with $L_1$ lying between the primaries.The remaining Lagrangian points $L_4$ and $L_5$ lie opposite sides of the joining the primaries. The three dimensional periodic orbits around Lagrangian points are called halo orbits. Halo orbits were introduced by  \cite{Farquhar1968}. He discovered the trajectories around the Earth-Moon $L_2$ which could be used to place a communication satellite that would continuously link between the Earth and the Moon. Other researchers \citep{Breakwell,Howell,Howell1984} have studied halo orbit families for the Earth-Moon system. ISEE-3 was the first halo orbit mission. A third order approximation was introduced by \cite{Richardson}	 to represent halo orbits in the Sun-Earth system.   
	 \cite{Tiwary2015} have computed a first guess of halo orbits upto fourth order approximation using the Lindstedt-Poincar$\acute{e}$ method in the photogravitational RTBP with oblateness.\\
	 In this paper we have computed halo orbits around the Lagrangian points $L_1$ and $L_2$ in the Sun-Earth system considering the Sun as radiating body and the Earth as oblate spheroid. The fourth order approximations to solutions given by \citep{Tiwary2015} have been improved incorporating fifth order approximation. The comparison of the orbits obtained by fourth and fifth order approximations are shown graphically. Equations of motion of an infinitesimal body in a synodic system are described in Section 2. Equations of motion around $L_1$ and $L_2$ and the procedure to obtain halo orbits are described in Section 3. Section 4 has discussion of effects of solar radiation pressure and oblateness on various parameters of halo orbits. In Section 5, the conclusions are given.   
	\noindent
	\section{Equations of Motion}
	Consider the photogravitational CRTBP with oblateness and assume that the bigger primary, the Sun, is the source of radiation and the smaller primary, the Earth, is an oblate spheroid. The mass reduction factor $q$, the perturbed mean motion $n$ and the oblateness coefficient $A_2$ are given as \citep{McCuskey,Sharma1987} $$ q = \left( 1 - \frac{F_p}{F_g}\right) , n = \sqrt{\left( 1 + \frac{3}{2}A_2\right)}, A_2 = \frac{R_e^2 - R_p^2}{5R^2}$$
	where $F_p$ is the solar radiation pressure force, $F_g$ is the gravitational attraction force, $R_e$ and $R_p$ are, respectively, the equatorial and polar radii of the smaller primary and $R$ is the distance between the two primaries. The expression for $q$ shows that as solar radiation pressure increases, $q$ decreases. We use synodic coordinates with origin at the centre of mass of the primaries for describing the motion of the system. Choose the unit of length as the distance between the primaries. Let $(x, y, z)$ be the coordinates of the infinitesimal body and $(-\mu, 0, 0)$ and $(1-\mu, 0, 0)$ be the coordinates of the bigger and smaller primaries, respectively, in the dimensionless synodic coordinate system, where $\mu= \case{m_2}{m_1+m_2}$, $m_1,m_2$ being the masses of the bigger and smaller  primaries. The equations of motion of infinitesimal body with oblateness and solar radiation pressure are given by \citep{Szebehely,Sharma1987,Tiwary2015}
	\begin{eqnarray}
		\ddot{x} - 2n\dot{y} &=& \frac{\partial\Omega^*}{\partial x},\label{eq:1}\\
		\ddot{y} + 2n\dot{x} &=& \frac{\partial\Omega^*}{\partial y},\label{eq:2}\\
		\ddot{z} &=& \frac{\partial\Omega^*}{\partial z},\label{eq:3}	
	\end{eqnarray}
	where
	\begin{equation}
	\Omega^* = n^2\frac{(x^2 + y^2)}{2} + \frac{(1 - \mu)q}{r_1} + \frac{\mu}{r_2} + \frac{\mu A_2}{2r_2^3},
	\end{equation}
	and	 
	 \begin{eqnarray}
	     r_1 &=& \sqrt{(x+\mu)^2 + y^2 + z^2} ,\\
	     r_2 &=& \sqrt{(x + \mu -1)^2 + y^2 + z^2} 	
	 \end{eqnarray}
	  are the distances of the infinitesimal body from the bigger and smaller primaries, respectively.
	
	 \section{Computation of Halo orbits}
	 Lindstedt-Poincar$\acute{e}$ method\citep{Koon} is used to compute the halo orbits around the libration points $L_1$ and $L_2$. 
	 It is used for solving non-linear ordinary differential equation when the regular perturbation method fails by removing secular terms and thereby converting to weakly non-linear equation with finite oscillatory solutions.
	 \subsection{Equations of motion near $L_1$ and $L_2$}
	 To obtain the halo orbits around the Lagrangian point the origin is shifted to the location of the Lagrangian point. Then the new coordinates are given by \citep{Koon}
	 \begin{eqnarray}
	 	X &=& \frac{1}{\gamma}(x + \mu -1 \pm \gamma),\label{eq:7}\\
	 	Y &=& \frac{1}{\gamma}y,\label{eq:8}\\
	 	Z &=& \frac{1}{\gamma}z\label{eq:9},
	 \end{eqnarray}
	 where $\gamma$ is the distance between the Lagrangian point and the smaller primary. In \eqref{eq:7},  upper sign corresponds to $L_1$ and lower sign corresponds to $L_2$.  The variables $X, Y$ and $Z$ are normalized so that the distance between the Lagrangian point and the smaller primary is $1$. Using the above transformation in the equations of motion \eqref{eq:1}-\eqref{eq:3}, we obtain
	 \begin{eqnarray}
		\gamma(\ddot{X} - 2n\dot{Y}) &=& \frac{1}{\gamma}\frac{\partial\Omega}{\partial X},\label{eq:10}\\
		\gamma(\ddot{Y} + 2n\dot{X}) &=& \frac{1}{\gamma}\frac{\partial\Omega}{\partial Y},\label{eq:11}\\
		\gamma\ddot{Z} &=& \frac{1}{\gamma}\frac{\partial\Omega}{\partial Z},\label{eq:12}
	 \end{eqnarray}
	 where
	 \begin{equation}\label{Omega}
	 \Omega = \frac{n^2}{2}[(\gamma X + 1 - \mu \mp \gamma)^2 + (\gamma Y)^2] + \frac{(1 - \mu)q}{R_1} + \frac{\mu}{R_2} + \frac{\mu A_2}{2R_2^3}
	 \end{equation}
	 and
	 \begin{eqnarray*}
	 	R_1 &=& \sqrt{(\gamma X + 1 \mp \gamma)^2 + (\gamma Y)^2+ (\gamma Z)^2} ,\\
	 	R_2 &=& \sqrt{(\gamma X \mp \gamma)^2 + (\gamma Y)^2 + (\gamma Z)^2}.
	 \end{eqnarray*}
	 Expanding the nonlinear terms $\frac{(1 - \mu)q}{R_1} + \frac{\mu}{R_2} + \frac{\mu A_2}{2R_2^3}$ of \eqref{Omega} using Legendre polynomials , equations of motion can be written as \citep{Koon,Tiwary2015}
	  \begin{eqnarray}
	  	\ddot{X} - 2n\dot{Y} - (n^2 +2C_2)X =\frac{\partial}{\partial X}\sum_{k\geqslant 3}C_k\rho^kP_k\left( \frac{X}{\rho}\right) ,\label{eq:14}\\
	  	\ddot{Y} + 2n\dot{X} + (C_2 - n^2)Y =\frac{\partial}{\partial Y}\sum_{k\geqslant 3}C_k\rho^kP_k\left( \frac{X}{\rho}\right) ,\label{eq:15}\\
	  	\ddot{Z} + C_2Z =\frac{\partial}{\partial Z}\sum_{k\geqslant 3}C_k\rho^kP_k\left( \frac{X}{\rho}\right) \label{eq:16}.
	  \end{eqnarray}
	  In above equations, the left hand side contains the linear terms and the right hand side contains the non-linear terms. The coefficients $C_k$ are given by
	  \begin{equation}\label{key}
	  C_k = \frac{1}{\gamma ^3}\hskip-3pt\left[ \frac{(-1)^kq(1 - \mu)\gamma^{k+1}}{(1 \mp \gamma)^{k+1}}\,  +(\pm 1)^k\left(  \mu \, + \frac{3\mu A_2}{2\gamma^2}\right)  \right] , 
	  \end{equation}
	  for $k\geqslant 1$.
	  Considering only linear terms in equations \eqref{eq:14}-\eqref{eq:16}, the solution of the linearized equations is
\begin{equation*}
      X(t) = A_1e^{\alpha t} + A_2e^{-\alpha t} + A_3\cos\lambda t + A_4\sin\lambda t,
\end{equation*}
\vspace{-20pt}
\begin{dmath*}
	Y(t) = -k_1A_1e^{\alpha t} + k_1A_2e^{-\alpha t} - k_2A_3\sin\lambda t +k_2A_4\cos\lambda t,
\end{dmath*}
\vspace{-10pt}
\begin{equation*}
      Z(t) = A_5\cos\sqrt{C_2}t + A_6\sin\sqrt{C_2}t,
\end{equation*}
  where $A_1, A_2, A_3, A_4, A_5$ and $A_6$ are arbitrary constants,
  \begin{align*}
  \alpha =&  \sqrt{\frac{-(2n^2-C_2) + \sqrt{9C_2^2-8n^2C_2}}{2}},\\
  \lambda =& \sqrt{\frac{2n^2-C_2+ \sqrt{9C_2^2-8n^2C_2}}{2}},\\
  k_1 =& \frac{(2C_2+n^2)-\alpha^2}{2n\alpha},\\
  k_2 =& \frac{(2C_2+n^2)+\lambda^2}{2n\lambda}.
  \end{align*}
  Linearized equations corresponding to equations \eqref{eq:14}-\eqref{eq:16} have two real roots which are equal in magnitude and opposite in sign. If the initial conditions are chosen arbitrarily, then these roots give rise to unbounded solutions.To avoid this, we take $A_1 = A_2 = 0$ and $A_3 = -A_X\cos\phi, A_4 = A_X\sin\phi, A_5 = A_Z\sin\psi$ and $A_6 = A_Z\cos\psi$ and get the bounded solution in the following form: \citep{Koon}
  \begin{align*}
  X(t) =& -A_X\cos(\lambda t + \phi),\\
  Y(t) =& kA_X\sin(\lambda t + \phi),\\
  Z(t) =& A_Z\sin(\sqrt{C_2}t + \psi),
  \end{align*}
  where $A_X$ and $A_Z$ are amplitudes; $\lambda$ and $\sqrt{C_2}$ are the frequencies; $k = k_2$; $\phi$ and $\psi$ are phases of the in-plane and out of plane motions respectively. The ratio of $\lambda$ and $\sqrt{C_2}$ is irrational. This gives Lissajous(quasi periodic) orbits.

  \subsection{Lindstedt-Poincar$\acute{e}$ Method for the Halo Orbits}
      Halo orbits are important for spacecraft mission design. Many researchers have obtained the halo orbits upto third order approximation\citep{Richardson,Howell1984,Breakwell,Koon,Prithiviraj&Sharma2016,Pushparaj&Sharma2016,Saurav&Sharma2019}. \cite{Tiwary2015} have computed halo orbits upto fourth order approximation with the Sun as a radiating body and the Earth as an oblate spheroid using Lindstedt-Poincar$\acute{e}$ method. Here, we have computed halo orbits upto fifth order approximation with radiation pressure and oblateness using Lindstedt-Poincar$\acute{e}$ method. The non-linear terms in \eqref{eq:14}-\eqref{eq:16} change the frequency of the linearized system. Due to this secular terms appear in successive approximations. To change the frequency, we take a new independent variable $\tau = \omega t$, where $\omega$ is a frequency connection. Then the equations of motion \eqref{eq:14}-\eqref{eq:16} in terms of $\tau$ truncated at degree $5$ are:
   \begin{multline}\label{eq:18}
   \omega^2X'' - 2n\omega Y' - (n^2 + 2C_2)X \\= \frac{3}{2}C_3(2X^2-Y^2-Z^2)+2C_4X(2X^2-3Y^2-3Z^2)\\\quad+\frac{5}{8}C_5[8X^2\{X^2-3(Y^2+Z^2)\}+3(Y^2+Z^2)^2]\\\quad+3C_6[2X^3\{X^2-5(Y^2+Z^2)\}+\frac{15}{4}X(Y^2+Z^2)^2],
   \end{multline}
   \begin{multline}\label{eq:19}
   	\omega^2Y'' + 2n\omega X' +(C_2-n^2)Y\\ = -3C_3XY-\frac{3}{2}C_4Y(4X^2-Y^2-Z^2)\\\quad-\frac{5}{2}C_5XY(4X^2-3Y^2-3Z^2)\\\quad+\frac{15}{2}C_6[X^2Y\{-2X^2+3(Y^2+Z^2)\}-\frac{1}{4}Y(Y^2+Z^2)^2],
   \end{multline}
   \begin{multline}\label{eq:20}
   	\omega^2 Z''+ \lambda^2Z \\= -3C_3XZ -\frac{3}{2}C_4Z(4X^2-Y^2-Z^2)\\\quad-\frac{5}{2}C_5XZ(4X^2-3Y^2-3Z^2)\\\quad+\frac{15}{2}C_6[X^2Z\{-2X^2+3(Y^2+Z^2)\}-\frac{1}{4}Z(Y^2+Z^2)^2]\\+ \Delta Z,
   	\qquad\qquad\qquad\qquad\qquad\qquad\qquad\qquad\,\,\,\,\,\,
   \end{multline}
   where $\Delta = \lambda^2 - C_2$ is the frequency correction term to obtain halo orbit and $\Delta = \mathcal{O}(\epsilon^2)$.\\
   The solutions of  \eqref{eq:18}-\eqref{eq:20} are assumed in the perturbations form as \citep{Thurman}:
   \begin{align}
   X(\tau)=& \epsilon X_1(\tau) + \epsilon^2X_2(\tau) + \epsilon^3X_3(\tau) + \epsilon^4X_4(\tau)\nonumber\\& + \epsilon^5X_5(\tau) + . . . \, ,\label{eq:21}\\
   Y(\tau) =& \epsilon Y_1(\tau) + \epsilon^2Y_2(\tau) + \epsilon^3Y_3(\tau) + \epsilon^4Y_4(\tau)\nonumber\\& + \epsilon^5Y_5(\tau) + . . . \, ,\label{eq:22}\\
   Z(\tau) =& \epsilon Z_1(\tau) + \epsilon^2Z_2(\tau) + \epsilon^3Z_3(\tau) + \epsilon^4Z_4(\tau)\nonumber\\& + \epsilon^5Z_5(\tau) + . . . ,\label{eq:23}
   \end{align}
   and let
   \begin{equation}\label{eq:24}
   \omega = 1+\epsilon\omega_1 + \epsilon^2\omega_2 + \epsilon^3\omega_3 + \epsilon^4\omega_4 + . . .
   \end{equation}
   Substituting the solutions \eqref{eq:21}-\eqref{eq:24} into equations of motion \eqref{eq:18}-\eqref{eq:20} and equating the coefficients of the same order of $\epsilon, \epsilon^2, \epsilon^3,$ and $\epsilon^4$, we obtain the first, second, third and fourth order equations, respectively \citep{Thurman,Tiwary2015}. For obtaining more accurate solutions of the equations we have collected the coefficients of $\epsilon^5$ and obtained the fifth order equations.

   \subsubsection{Fifth Order Equations}
   Collecting the coefficients of $\epsilon^5$ and incorporating all the solutions and conditions used upto fourth order approximations\citep{Tiwary2015}, we get the fifth order equations as :
     \begin{align}
     X_5'' - 2nY_5' -(n^2 + 2C_2)X_5 = \gamma_{51}\label{eq:25}\\
     Y_5'' + 2nX_5' + (C_2 - n^2)Y_5 = \gamma_{52}\,\,\,\label{eq:26}\\
     Z_5'' + \lambda^2Z_5 = \left\lbrace \begin{array}{c}
     f_3, \quad p = 0, 2 \\
     f_4, \quad p = 1, 3
     \end{array}\right.\quad\label{eq:27}
   \end{align}
   where
   \begin{dmath*}
    \gamma_{51} = [v_4 + 2\lambda A_X\omega_4(nk-\lambda)]\cos\tau_1 + \gamma_8\cos3\tau_1+ \gamma_9\cos5\tau_1,
\end{dmath*}
\vspace{-10pt}
\begin{dmath*}
	\gamma_{52} =[v_5+ 2\lambda A_X\omega_4(\lambda k -n)]\sin\tau_1 + \beta_9\sin3\tau_1 + \beta_{10}\sin5\tau_1,
\end{dmath*}
\vspace{-10pt}
   \begin{dmath*}
   	f_3 =[v_6 \pm 2\omega_4\lambda^2 A_Z]\sin\tau_1 + \delta_8\sin3\tau_1 + \delta_9\sin5\tau_1,
   \end{dmath*} 
\vspace{-10pt}
    \begin{dmath*}
    	f_4 = [v_6 \pm 2\omega_4\lambda^2 A_Z]\cos\tau_1 + \delta_8\cos3\tau_1 + \delta_9\cos5\tau_1
    \end{dmath*}    
   and the remaining coefficients are given in Appendix.
   In $f_3$, upper sign corresponds to $p = 0$ and lower sign corresponds to $p = 2$. Similarly, in $f_4$,  upper sign corresponds to $p = 1$ and lower corresponds to $p=3$.\\
   The secular term can be removed from \eqref{eq:27} if
   \begin{equation}\label{eq:28}
   v_6 \pm 2\omega_4\lambda^2A_Z = 0,
   \end{equation}
   where the upper sign corresponds to $p = 0,1$ and the lower sign corresponds to $p = 2,3$.\\

   \noindent
   To remove the secular terms from \eqref{eq:25} and \eqref{eq:26}, we use a single condition from their particular solution \citep{Thurman,Tiwary2015}
   \begin{equation}\label{eq:29}
   [v_4 + 2\lambda A_X\omega_4(nk-\lambda)] - k[v_5+ 2\lambda A_X\omega_4(\lambda k -n)] = 0.
   \end{equation}
   From equation \eqref{eq:29}, we get
   \begin{equation}\label{eq:30}
     \omega_4 = \dfrac{v_4 - kv_5}{2\lambda A_x(\lambda(k^2 + 1)-2nk)}.
   \end{equation}
   Using conditions \eqref{eq:28} and \eqref{eq:30} in equations \eqref{eq:25}-\eqref{eq:27}, the equations of motion take the following form:
   \begin{multline}\label{eq:31}
   	X_5'' - 2nY_5' - (n^2 + 2C_2)X_5 \\= k\beta_{11}\cos\tau_1 + \gamma_8\cos3\tau_1 + \gamma_9\cos5\tau_1,\qquad\qquad
   \end{multline}
   \begin{multline}\label{eq:32}
   	Y_5'' + 2nX_5' + (C_2 - n^2)Y_5 \\= \beta_{11}\sin\tau_1 + \beta_9\sin3\tau_1 + \beta_{10}\sin5\tau_1,\qquad\qquad
   \end{multline}
   \begin{multline}\label{eq:33}
   	Z_5'' + \lambda^2Z_5 = \left\lbrace \begin{array}{c}
    \delta_8\sin3\tau_1 + \delta_9\sin5\tau_1, \quad p = 0, 2, \\
   	\delta_8\cos3\tau_1 + \delta_9\cos5\tau_1, \quad p = 1, 3,
   	\end{array}\right.
   \end{multline}
   where $\beta_{11} = v_5 + 2\lambda A_X\omega_4(\lambda k -n)$. The solution of equations \eqref{eq:31}-\eqref{eq:33} is given by
   \begin{align}
   X_5(\tau) =& \rho_{51}\cos3\tau_1 + \rho_{52}\cos5\tau_1,\label{eq:34}\\
   Y_5(\tau) =& \sigma_{51}\sin\tau_1 + \sigma_{52}\sin3\tau_1 + \sigma_{53}\sin5\tau_1,\label{eq:35}\\
   Z_5(\tau) =& \left\lbrace \begin{array}{c}
   k_{51}\sin3\tau_1 + k_{52}\sin5\tau_1, \quad p = 0,2, \\
   k_{51}\cos3\tau_1 + k_{52}\cos5\tau_1, \quad p = 1,3,
   \end{array}\right.\label{eq:36}
   \end{align}
   where the coefficients are given in the Appendix.

   \subsubsection{Final Approximation}
   Final approximation is obtained  by removing $\epsilon$ from all the equations. For that we take the mapping $A_X \rightarrow \frac{A_X}{\epsilon}$ and $A_Z \rightarrow \frac{A_Z}{\epsilon}$. Combining the solutions component wise in \eqref{eq:21}-\eqref{eq:23}, we get \citep{Tiwary2015}
   \begin{align}
   X(\tau) = &(\rho_{20} + \rho_{40}) -A_X\cos\tau_1\nonumber\\& + (\rho_{21}+ \zeta\rho_{22}+\rho_{41})\cos2\tau_1\nonumber\\& + (\rho_{31}+\rho_{51})\cos3\tau_1 + \rho_{42}\cos4\tau_1 + \rho_{52}\cos5\tau_1,\label{eq:37}\\
   Y(\tau)= &(kA_X + \sigma_{32} + \sigma_{51})\sin\tau_1 \nonumber\\&+ (\sigma_{21}+\sigma_{41}+\zeta\sigma_{22})\sin2\tau_1 \nonumber\\& + (\sigma_{31} + \sigma_{52})\sin3\tau_1+\sigma_{42}\sin4\tau_1+\sigma_{53}\sin5\tau_1 ,\label{eq:38}\\
   Z(\tau) =& \left\lbrace \begin{array}{c}
   f_5, \quad p = 0, 2 \\
   f_6, \quad p = 1, 3
   \end{array}\right.\label{eq:39}
   \end{align}
   where
\begin{dmath*}
	 f_5 = (-1)^{\frac{p}{2}}(A_Z\sin\tau_1 +k_{21}\sin2\tau_1 + k_{31}\sin3\tau_1) + k_{41}\sin2\tau_1 + k_{42}\sin4\tau_1+k_{51}\sin3\tau_1 + k_{52}\sin5\tau_1,
\end{dmath*}
\vspace{-10pt}
\begin{dmath*}	 
	f_6 =(-1)^{\frac{p-1}{2}}(A_Z\cos\tau_1 +k_{21}\cos2\tau_1 +k_{22}+ k_{32}\cos3\tau_1) +k_{40}+ k_{41}\cos2\tau_1 + k_{42}\cos4\tau_1+k_{51}\cos3\tau_1 + k_{52}\cos5\tau_1.
\end{dmath*}
   Using equations \eqref{eq:37}-\eqref{eq:39}, we can get the first guess of halo orbits.
   
   \section{Discussion}
   The halo orbits in the photogravitational Sun-Earth system with oblateness upto fourth order approximations using Lindstedt-Poincar$\acute{e}$ method are obtained by \cite{Tiwary2015}. Here, the first guess of the halo orbit in the same system is obtained upto fifth order approximation using Lindstedt-Poincar$\acute{e}$ method. Equations \eqref{eq:37}-\eqref{eq:39} are used with the amplitudes $A_X = 206000$ km and $A_Z = 110000$ km from the ISEE-$3$ mission.
   
\begin{figure}
	\includegraphics[width=2.8in]{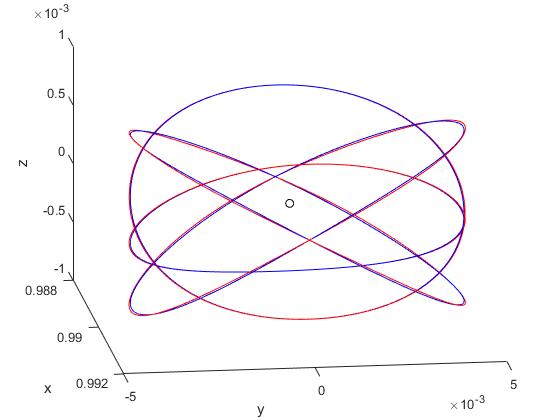}	
	\caption{$4^{th}$ and $5^{th}$ order halo orbits around $L_1$ corresponding to $A_2 = 2.4\times 10^{-12}, q = 0.9995$}
	\label{fig:1}
\end{figure}

\begin{figure}
	\includegraphics[width=2.8in]{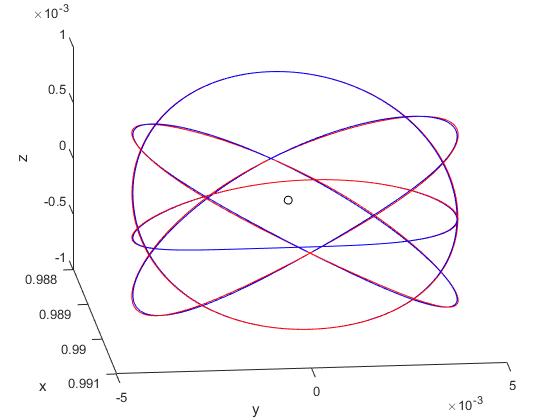}
	\caption{$4^{th}$ and $5^{th}$ order halo orbits around $L_1$ corresponding to $A_2 = 2.4\times 10^{-12}, q = 0.9945$}
	\label{fig:2}
\end{figure}

\begin{figure}
	\includegraphics[width=2.8in]{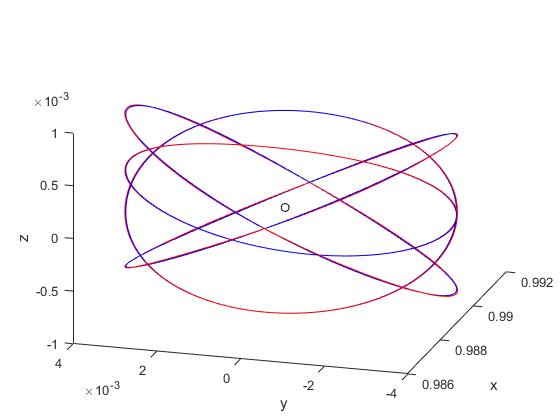}
	\caption{$4^{th}$ and $5^{th}$ order halo orbits around $L_1$ corresponding to $A_2 = 2.4\times 10^{-12}, q = 0.9895$}
	\label{fig:3}
\end{figure}

\begin{figure}
	\includegraphics[width=2.8in]{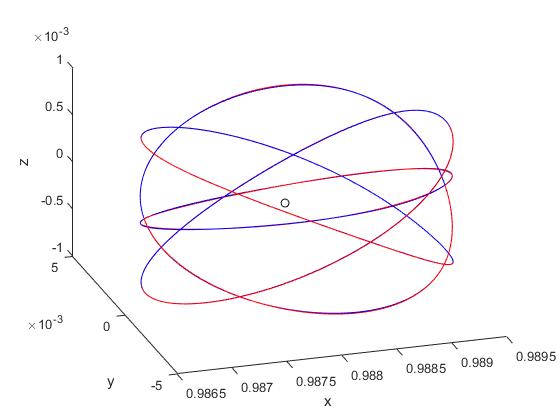}
	\caption{$4^{th}$ and $5^{th}$ order halo orbits around $L_1$ corresponding to $A_2 = 2.4\times 10^{-12}, q = 0.9845$}
	\label{fig:4}
\end{figure}

 \placefigure{fig:1}
\placefigure{fig:2}
   \placefigure{fig:3}
   \placefigure{fig:4}  
   \noindent
   The orbits are plotted for different values of phases. Fig.$\ref{fig:1}$ to Fig.$\ref{fig:4}$ show halo orbits around $L_1$ for different values of $q$, mass reduction factor.
   Orbits coloured in blue represents fourth order orbits and red corresponds to fifth order orbits. 
%

\begin{figure}
	\includegraphics[width=2.8in]{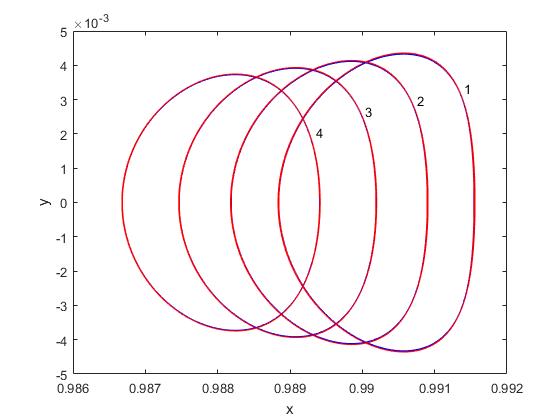}
	\caption{Effect of radiation pressure on the position of halo orbits around $L_1$}
	\label{fig:6}
\end{figure}

\placefigure{fig:6}

\begin{figure}
	\includegraphics[width=2.8in]{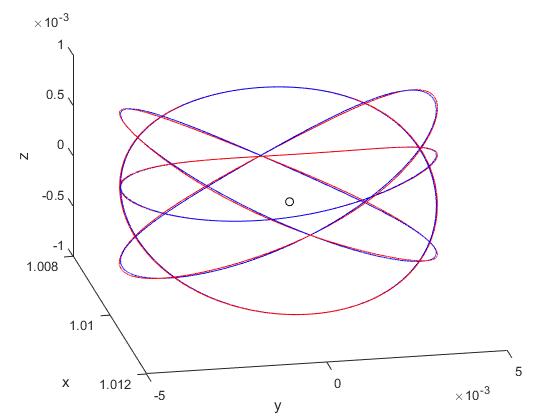}
	\caption{$4^{th}$ and $5^{th}$ order halo orbits around $L_2$ corresponding to $A_2 = 2.4\times 10^{-12}, q = 0.9995$}
	\label{fig:8}
\end{figure}

\begin{figure}
	\includegraphics[width=2.8in]{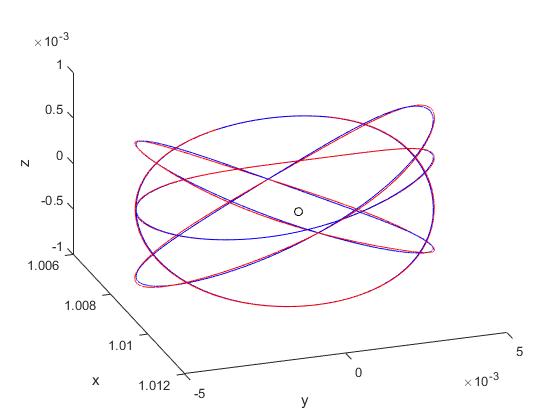}
	\caption{$4^{th}$ and $5^{th}$ order halo orbits around $L_2$ corresponding to $A_2 = 2.4\times 10^{-12}, q = 0.9945$}
	\label{fig:9}
\end{figure}

\begin{figure}
	\includegraphics[width=2.8in]{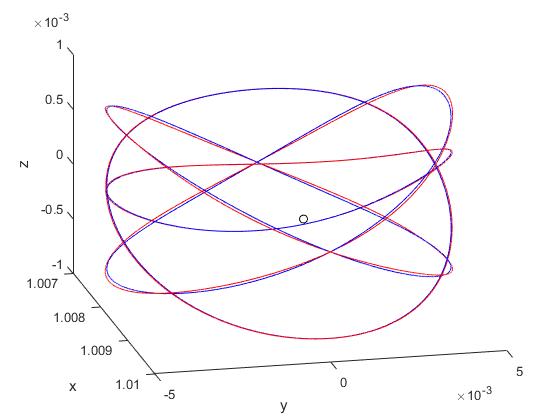}
	\caption{$4^{th}$ and $5^{th}$ order halo orbits around $L_2$ corresponding to $A_2 = 2.4\times 10^{-12}, q = 0.9895$}
	\label{fig:10}
\end{figure}

\begin{figure}
	\includegraphics[width=2.8in]{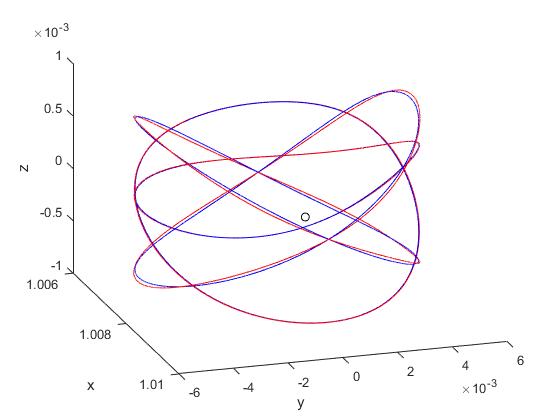}
	\caption{$4^{th}$ and $5^{th}$ order halo orbits around $L_2$ corresponding to $A_2 = 2.4\times 10^{-12}, q = 0.9845$}
	\label{fig:11}
\end{figure}

\noindent
The effects of radiation pressure on the position of halo orbits are given in Fig.$\ref{fig:6}$. Fig.$\ref{fig:6}$ shows the positions of halo orbits for $q = 0.9995,0.9945,0.9895$ and $0.9845$ labeled as $1,2,3$ and $4$, respectively, with 
the actual oblateness of Earth. As radiation pressure increases, the halo orbits move towards the source of radiation. This agrees with conclusions of  \cite{Eapen2014}.
\placefigure{fig:8}
\placefigure{fig:9}
\placefigure{fig:10}
\placefigure{fig:11}

\begin{figure}
	\includegraphics[width=2.8in]{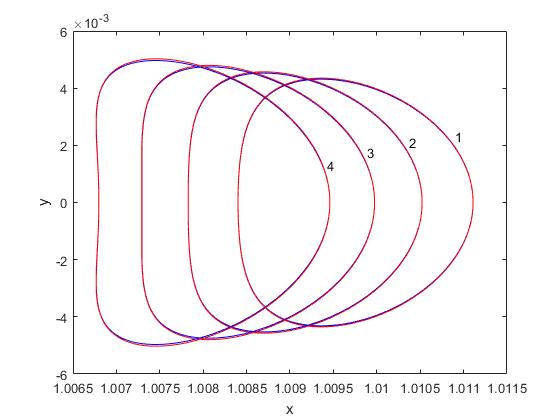}
	\caption{Effect of radiation pressure on the position of halo orbits around $L_2$}
	\label{fig:13}
\end{figure}

\begin{figure}
	\includegraphics[width=2.8in]{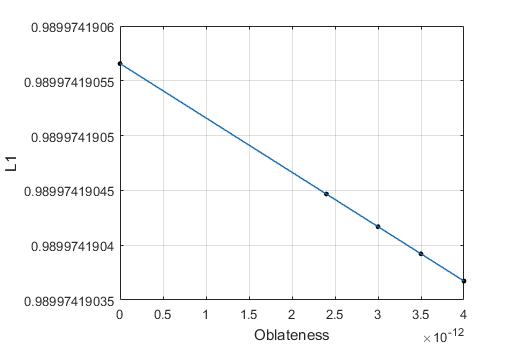}
	\caption{Effect of oblateness on the position of $L_1$}
	\label{fig:131}
\end{figure}

\begin{figure}
	\includegraphics[width=2.8in]{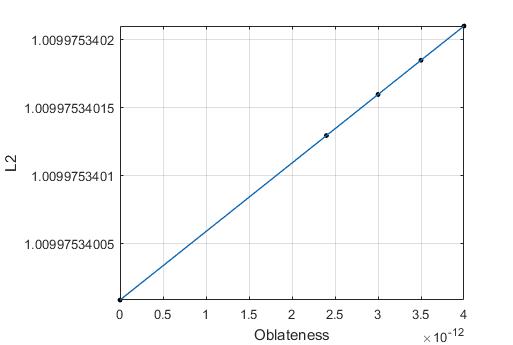}
	\caption{Effect of oblateness on the position of $L_2$}
	\label{fig:14}
\end{figure}

\begin{figure}
	\includegraphics[width=2.8in]{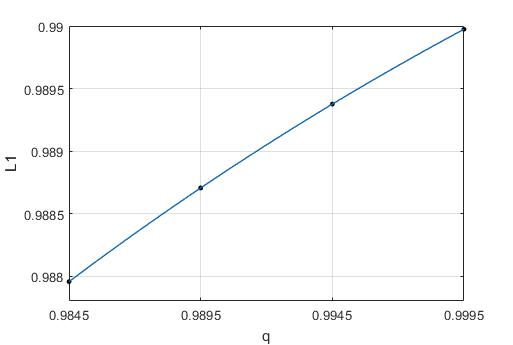}
	\caption{Effect of radiation pressure on the position of $L_1$}
	\label{fig:15}
\end{figure}

\begin{figure}
	\includegraphics[width=2.8in]{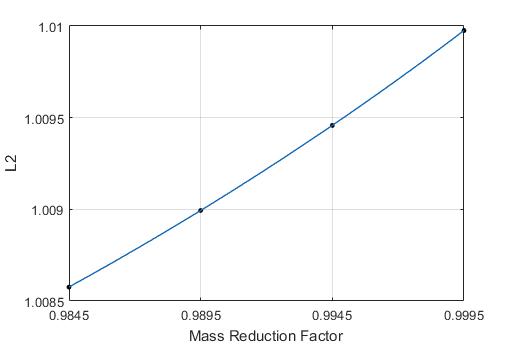}
	\caption{Effect of radiation pressure on the position of $L_2$}
	\label{fig:16}
\end{figure}

\begin{figure}
	\includegraphics[width=2.8in]{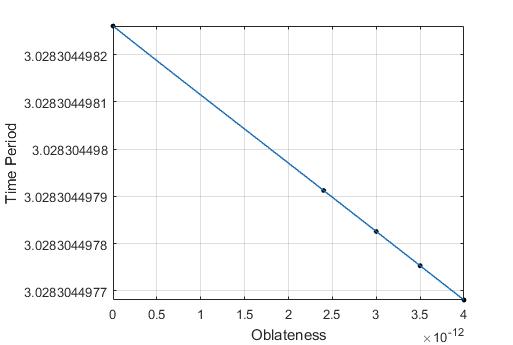}
	\caption{Effect of oblateness on time period of halo orbits around $L_1$}
	\label{fig:17}
\end{figure}

\begin{figure}
	\includegraphics[width=2.8in]{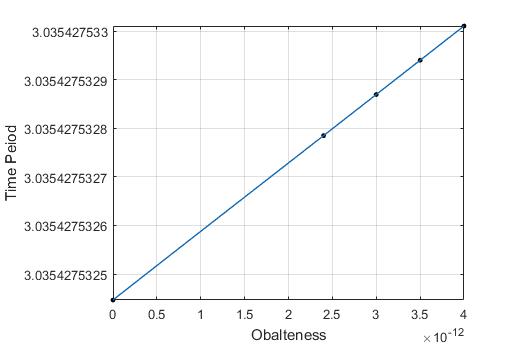}
	\caption{Effect of oblateness on time period of halo orbits around $L_2$}
	\label{fig:18}
\end{figure}

\begin{figure}
	\includegraphics[width=2.8in]{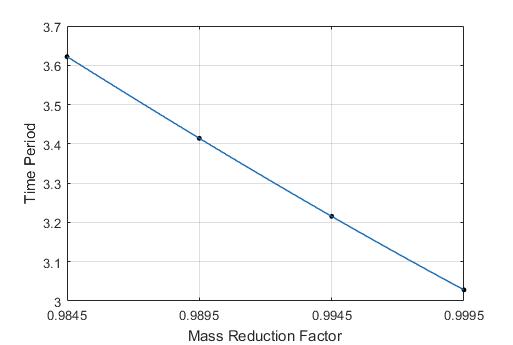}
	\caption{Effect of radiation pressure on time period of halo orbits around $L_1$}
	\label{fig:19}
\end{figure}

\begin{figure}
	\includegraphics[width=2.8in]{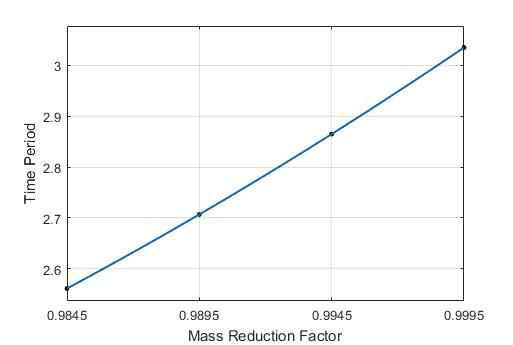}
	\caption{Effect of radiation pressure on time period of halo orbits around $L_2$}
	\label{fig:20}
\end{figure}

\noindent
Fig.$\ref{fig:8}-\ref{fig:11}$ represent halo orbits around $L_2$ corresponding to mass reduction factor $0.9995,0.9945,0.9895$ and $0.9845$, respectively, with oblateness $A_2 = 2.4\times 10^{-12}$, the obalteness of the Earth.  
\placefigure{fig:13}
\noindent
 Fig.$\ref{fig:13}$ shows the variation in position of halo orbits due to radiation pressure. Here, the orbits labeled as $1,2,3$ and $4$ correspond to $q = 0.9995,0.9945,0.9895,0.9845$ and oblateness is  $2.4\times 10^{-12}$. Halo orbits move towards the source of radiation with the increase in radiation pressure. 
\placetable{tbl:1}
\placetable{tbl:2}
\placetable{tbl:3}
\placetable{tbl:4}
\noindent
Table$\ref{tbl:1}$ , Table$\ref{tbl:2}$, Table$\ref{tbl:3}$ and Table$\ref{tbl:4}$ show the variation in coefficients, the position of Lagrangian points, $\Delta$ and time period $\tau$ due to variation in radiation pressure and oblateness. $\Delta = \lambda^2 - C_2$ is the frequency correction term to obtain the halo orbits. $\tau$ is the time taken by the infinitesimal body to complete one rotation about the Lagrangian point. 
 Table$\ref{tbl:1}$ shows the effect of radiation pressure on parameters of orbits around $L_1$. It can be observed that as the radiation pressure increases, that is, $q$ decreases, $L_1$ move towards the source of radiation, the Sun. Also, the time period of orbits increase with the increase in radiation pressure. Table$\ref{tbl:2}$ represents the effect of oblateness on various parameters of orbits around $L_1$. With the increase in radiation pressure, orbits move towards the Sun and their time period is decreased. From Table$\ref{tbl:3}$, it can be observed that due to increase in radiation pressure, the orbits around $L_2$ move towards the Sun and their time period is decreased. Effect of oblateness on position of orbits and time period can be observed from Table $\ref{tbl:4}$. Halo orbits around $L_2$ move away from the source of radiation and also time period of orbits increase due to increase in oblateness.  
Fig.$\ref{fig:131}$ represents the effect of oblateness on position of $L_1$. As oblateness increases, $L_1$ moves towards the source of radiation, the Sun. In Fig.$\ref{fig:14}$, the reverse effect of oblateness is observed on the position of $L_2$. That is, as oblateness increases, $L_2$ moves away from the Sun. 
Fig.$\ref{fig:15}$ and Fig.$\ref{fig:16}$ show the variation in position of $L_1$ and $L_2$ due to radiation pressure, respectively. With the increase in radiation pressure, $L_1$ and $L_2$ both move towards the Sun.
The effect of radiation pressure and oblateness on time period is graphically shown in Fig.$\ref{fig:17}-\ref{fig:20}$. Time periods of halo orbits decrease with the increase in oblateness  around $L_1$ while they increase with the increase in oblateness  around $L_2$. With the increase in radiation pressure, time period of orbits around $L_1$ increases and decreases around $L_2$. 

\section{Conclusion}
Photogravitational RTBP with oblateness, where the Sun is radiating and the Earth an oblate spheroid, is studied for halo orbits. We have improved the fourth order equations obtained by \citep{Tiwary2015} using Lindstedt-Poincar$\acute{e}$ method to fifth order and to obtain halo orbits around $L_1$ and $L_2$. The deviations of the orbits around $L_1$ and $L_2$ obtained from fourth order and fifth order equations are shown graphically. The variations in position and time of halo orbits around $L_1$ and $L_2$ due to radiation pressure and oblateness are studied. It is found that the halo orbits around $L_1$ shift towards the source of radiation (Sun) as the radiation pressure and oblateness increase. However, the time period of halo orbits around $L_1$ increases with the increase in radiation pressure but decreases with the increase in the oblateness. Halo orbits around $L_2$ approaches the source of radiation with increase in the radiation pressure but recedes from the source of radiation due to increase in the oblateness. The period of halo orbits around $L_2$ decreases with increase in the radiation pressure but increases with increase in oblateness.  
\\

\acknowledgments\\
One of the authors(DS) would like to thank Council of Scientific and Industrial Research (CSIR) for financial support through JRF(File No. 09/114(0218)/2019-EMR-I).

\section*{Compliance with Ethical Standards}
\textbf{Conflict of Interest :} Author Dhwani Sheth has received  Junior Research Fellowship(JRF) from CSIR (File No. 09/114(0218)/2019-EMR-I).
\section*{Appendix}
\begin{align*}
v_4 = \left\lbrace 
\begin{array}{c}
v_{41}, \quad\text{when}\,\,p=0,2,\\
v_{42}, \quad\text{when}\,\,p=1,3.
\end{array}
\right. 
\end{align*}		
\begin{dmath*}
	v_{41} =
	\lambda\omega_2(2n\sigma_{32}-\lambda\omega_2A_X)+\frac{3}{2}C_3(-2A_X(2\rho_{40}+\rho_{41})+2\rho_{31}(\rho_{21}+\rho_{22})-kA_X\sigma_{41}
	-(\sigma_{21}+\sigma_{22})(\sigma_{31}+\sigma_{32})-(-1)^{\frac{p}{2}}A_Zk_{41}-k_{21}k_{31})+\frac{3}{2}C_4(2A_X^2\rho_{31}-4A_X((\rho_{20}+\rho_{21}
	+\rho_{22})^2+\rho_{20}^2)+2kA_X^2(\sigma_{31}+\sigma_{32})+2A_X(\sigma_{21}+\sigma_{22})^2
	-4kA_X\rho_{20}(\sigma_{21}+\sigma_{22})+k^2A_X^2\rho_{31}+2A_XA_Zk_{31}+2A_Xk_{21}^2
	-4A_Zk_{21}\rho_{20}+A_Z^2\rho_{31})+
	\frac{5}{2}C_5(-2A_X^3(3\rho_{20}+2(\rho_{21}+\rho_{22}))
	-3kA_X^3(\sigma_{21}+\sigma_{22})+3k^2A_X^3\rho_{20}-3A_X^2A_Zk_{21}
	+3A_XA_Z^2\rho_{20}+\frac{3}{4}k^3A_X^3(\sigma_{21}+\sigma_{22})
	+\frac{3}{4}k^2A_X^2A_Zk_{21}+\frac{3}{4}kA_XA_Z^2(\sigma_{21}
	+\sigma_{22})+\frac{3}{4}A_Z^3k_{21})+\frac{15}{32}C_6(-8A_X^5+8k^2A_X^5
	+8A_X^3A_Z^2-3k^4A_X^5-6k^2A_X^3A_Z^2-3A_XA_Z^4),	
\end{dmath*}
\begin{dmath*}
	v_{42} =\lambda\omega_2(2n\sigma_{32}-\lambda\omega_2A_X)
	+\frac{3}{2}C_3(-2A_X(2\rho_{40}+\rho_{41})+2\rho_{31}(\rho_{21}-\rho_{22})-kA_X\sigma_{41}
	-(\sigma_{21}-\sigma_{22})(\sigma_{31}+\sigma_{32})-(-1)^{\frac{p-1}{2}}A_Z(2k_{40}+k_{41})-k_{21}k_{32})+\frac{3}{2}C_4(2A_X^2\rho_{31}-4A_X((\rho_{20}+\rho_{21}
	-\rho_{22})^2+\rho_{20}^2)+2kA_X^2(\sigma_{31}+\sigma_{32})+2A_X(\sigma_{21}-\sigma_{22})^2
	-4kA_X\rho_{20}(\sigma_{21}-\sigma_{22})+k^2A_X^2\rho_{31}+2A_XA_Zk_{32}+2A_X(k_{21}^2+2k_{21}k_{22}+2k_{22}^2)
	-4A_Z(k_{21}+k_{22})(\rho_{20}+\rho_{21}-\rho_{22})-A_Z^2\rho_{31})+\frac{5}{2}C_5(-2A_X^3(3\rho_{20}
	+2(\rho_{21}-\rho_{22}))-3kA_X^3(\sigma_{21}-\sigma_{22})+3k^2A_X^3\rho_{20}
	-3A_X^2A_Z(2k_{21}+3k_{22})+3A_XA_Z^2(3\rho_{20}+2(\rho_{21}-\rho_{22}))
	+\frac{3}{4}k^3A_X^3(\sigma_{21}-\sigma_{22})+\frac{3}{4}k^2A_X^2A_Zk_{22}
	+\frac{3}{4}kA_XA_Z^2(\sigma_{21}-\sigma_{22})+\frac{3}{4}A_Z^3(2k_{21}+3k_{22}))
	+\frac{15}{32}C_6(-8A_X^5+8k^2A_X^5+40A_X^3A_Z^2
	-3k^4A_X^5-6k^2A_X^3A_Z^2-15A_XA_Z^4).
\end{dmath*}
\begin{align*}
v_5 = \left\lbrace 
\begin{array}{c}
v_{51}, \quad\text{when}\,\,p=0,2,\\
v_{52}, \quad\text{when}\,\,p=1,3.
\end{array}
\right. 
\end{align*}		
\begin{dmath*}
	v_{51} =\omega_2\lambda^2(2\sigma_{32}+\omega_2kA_X)-\frac{3}{2}C_3(-A_X\sigma_{41}
	+2\rho_{20}\sigma_{32}+(\rho_{21}+\rho_{22})(\sigma_{31}-\sigma_{32})-\rho_{31}(\sigma_{21}+\sigma_{22})
	+kA_X(2\rho_{40}-\rho_{41}))-\frac{3}{8}C_4(4A_X^2(\sigma_{31}+\sigma_{32})
	-16A_X\rho_{20}(\sigma_{21}+\sigma_{22})+8kA_X^2\rho_{31}+8kA_X(2\rho_{20}^2-2\rho_{20}(\rho_{21}+\rho_{22})
	+(\rho_{21}+\rho_{22})^2)+3k^2A_X^2(\sigma_{31}-3\sigma_{32})-6kA_X(\sigma_{21}
	+\sigma_{22})^2+2kA_XA_Zk_{31}-2kA_Xk_{21}^2-4A_Zk_{21}(\sigma_{21}+\sigma_{22})
	+A_Z^2(\sigma_{31}-3\sigma_{32}))+\frac{5}{8}C_5(4A_X^3(\sigma_{21}+\sigma_{22})
	+3k^3A_X^3(3\rho_{20}-2(\rho_{21}+\rho_{22}))-9k^2A_X^3(\sigma_{21}+\sigma_{22})
	-12kA_X^3\rho_{20}-6kA_X^2A_Zk_{21}-3A_XA_Z^2(\sigma_{21}+\sigma_{22})+3kA_XA_Z^2(3\rho_{20}-2(\rho_{21}+\rho_{22})))
	+\frac{15}{64}C_6(-8kA_X^5+12k^3A_X^5+12kA_X^3A_Z^2-5k^5A_X^5-10k^3A_X^3A_Z^2-5kA_XA_Z^4),
\end{dmath*}
\begin{dmath*}
	v_{52} =
	\omega_2\lambda^2(2\sigma_{32}+\omega_2kA_X)-\frac{3}{2}C_3(-A_X\sigma_{41}
	+2\rho_{20}\sigma_{32}+(\rho_{21}-\rho_{22})(\sigma_{31}-\sigma_{32})-\rho_{31}(\sigma_{21}-\sigma_{22})
	+kA_X(2\rho_{40}-\rho_{41}))-\frac{3}{8}C_4(4A_X^2(\sigma_{31}+\sigma_{32})
	-16A_X\rho_{20}(\sigma_{21}-\sigma_{22})+8kA_X^2\rho_{31}+8kA_X(2\rho_{20}^2-2\rho_{20}(\rho_{21}-\rho_{22})
	+(\rho_{21}-\rho_{22})^2)+3k^2A_X^2(\sigma_{31}-3\sigma_{32})-6kA_X(\sigma_{21}
	-\sigma_{22})^2+2kA_XA_Zk_{32}-2kA_X(k_{21}^2-2k_{21}k_{22}+2k_{22}^2)-4A_Zk_{22}(\sigma_{21}-\sigma_{22})
	-A_Z^2(\sigma_{31}+\sigma_{32}))+\frac{5}{8}C_5(4A_X^3(\sigma_{21}-\sigma_{22})
	+3k^3A_X^3(3\rho_{20}-2(\rho_{21}-\rho_{22}))-9k^2A_X^3(\sigma_{21}-\sigma_{22})
	-12kA_X^3\rho_{20}-6kA_X^2A_Zk_{22}-3A_XA_Z^2(\sigma_{21}-\sigma_{22})+3kA_XA_Z^2\rho_{20})
	+\frac{15}{64}C_6(-8kA_X^5+12k^3A_X^5+12kA_X^3A_Z^2-5k^5A_X^5-2k^3A_X^3A_Z^2-kA_XA_Z^4).
\end{dmath*}
\begin{align*}
v_6 = \left\lbrace 
             \begin{array}{c}
               v_{60}, \quad\text{when}\,\,p=0,\\
               v_{61}, \quad\text{when}\,\,p=1,\\
               v_{62},\quad\text{when}\,\,p=2,\\
               v_{63},\quad\text{when} \,\,p=3.
             \end{array}
             \right. 
\end{align*}		
\begin{dmath*}
	v_{60} =	
		\omega_2^2\lambda^2A_Z-\frac{3}{2}C_3(-A_Xk_{41}+k_{31}(\rho_{21}+\rho_{22})
		-\rho_{31}k_{21}+A_Z(2\rho_{40}-\rho_{41}))-\frac{3}{2}C_4(A_X^2k_{31}
		-4A_X\rho_{20}k_{21}+2A_XA_Z\rho_{31}+2A_Z\{2\rho_{20}(\rho_{20}
		-\rho_{21}-\rho_{22})+ (\rho_{21} + \rho_{22})^2\}+\frac{1}{4}k^2A_X^2k_{31}
		-kA_Xk_{21}(\sigma_{21}+\sigma_{22})+\frac{1}{2}kA_XA_Z(\sigma_{31}-3\sigma_{32})
		+\frac{3}{4}A_Z^2k_{31}-\frac{1}{2}A_Z(\sigma_{21}+\sigma_{22})^2-\frac{3}{2}A_Zk_{21}^2)
		-\frac{5}{8}C_5(-4A_X^3k_{21}+12A_X^2A_Z\rho_{20}+9A_XA_Z^2k_{21}
		-3A_Z^3(3\rho_{20}-2(\rho_{21}+\rho_{22}))+3k^2A_X^3k_{21}
		-6kA_X^2A_Z(\sigma_{21}+\sigma_{22})-3k^2A_X^2A_Z\rho_{20})
		+\frac{15}{64}C_6(-8A_X^4A_Z+12k^2A_X^4A_Z+12A_X^2A_Z^3
		-5k^4A_X^4A_Z-10k^2A_X^2A_Z^3-5A_Z^5),
\end{dmath*}
\begin{dmath*}
	v_{61} = \omega_2^2\lambda^2A_Z-\frac{3}{2}C_3(-A_X(2k_{40}+k_{41})+k_{32}(\rho_{21}
	-\rho_{22})+\rho_{31}k_{21}+A_Z(2\rho_{40}+\rho_{41}))-\frac{3}{2}C_4(A_X^2
	 k_{32}-4A_X(\rho_{20}(k_{21}+2k_{22})+(\rho_{21}-\rho_{22})(k_{21}	
	+k_{22}))-2A_XA_Z\rho_{31}+2A_Z\{2\rho_{20}(\rho_{20}+\rho_{21}
	-\rho_{22})+ (\rho_{21} - \rho_{22})^2\}+\frac{1}{4}k^2A_X^2k_{32}-kA_X
	 k_{22}(\sigma_{21}-\sigma_{22})
	-\frac{1}{2}kA_XA_Z(\sigma_{31}+\sigma_{32})
	-\frac{3}{4}A_Z^2k_{32}-\frac{1}{2}A_Z(\sigma_{21}
	-\sigma_{22})^2-\frac{3}{2}A_Z(k_{21}^2
	+2k_{21}k_{22}+2k_{22}^2))-\frac{5}{8}C_5(-4A_X^3 (2k_{21}
	+3k_{22})+12A_X^2A_Z(3\rho_{20}+2(\rho_{21}-\rho_{22}))
	+9A_XA_Z^2(2k_{21}+3k_{22})-3A_Z^3(3\rho_{20}+2(\rho_{21}
	-\rho_{22}))         +3k^2A_X^3k_{22}+6kA_X^2A_Z(\sigma_{21}-\sigma_{22})
	-3k^2A_X^2A_Z\rho_{20})+\frac{15}{64}C_6(-40A_X^4A_Z
	+12k^2A_X^4A_Z
	+60A_X^2A_Z^3-k^4A_X^4A_Z
	-2k^2A_X^2A_Z^3-5A_Z^5),
\end{dmath*}
\begin{dmath*}
	v_{62}=
			-\omega_2^2\lambda^2A_Z-\frac{3}{2}C_3(-A_Xk_{41}-k_{31}(\rho_{21}+\rho_{22})
	+\rho_{31}k_{21}-A_Z(2\rho_{40}-\rho_{41}))-\frac{3}{2}C_4(-A_X^2k_{31}
	+4A_X\rho_{20}k_{21}-2A_XA_Z\rho_{31}-2A_Z\{2\rho_{20}(\rho_{20}
	-\rho_{21}-\rho_{22})+ (\rho_{21} + \rho_{22})^2\}-\frac{1}{4}k^2A_X^2k_{31}
	+kA_Xk_{21}(\sigma_{21}+\sigma_{22})-\frac{1}{2}kA_XA_Z(\sigma_{31}-3\sigma_{32})
	-\frac{3}{4}A_Z^2k_{31}+\frac{1}{2}A_Z(\sigma_{21}+\sigma_{22})^2+\frac{3}{2}A_Zk_{21}^2)
	-\frac{5}{8}C_5(4A_X^3k_{21}-12A_X^2A_Z\rho_{20}-9A_XA_Z^2k_{21}
	+3A_Z^3(3\rho_{20}-2(\rho_{21}+\rho_{22}))-3k^2A_X^3k_{21}
	+6kA_X^2A_Z(\sigma_{21}+\sigma_{22})+3k^2A_X^2A_Z\rho_{20})
	+\frac{15}{64}C_6(8A_X^4A_Z-12k^2A_X^4A_Z-12A_X^2A_Z^3
	+5k^4A_X^4A_Z+10k^2A_X^2A_Z^3+5A_Z^5),
\end{dmath*}
\begin{dmath*}
	v_{63} =
			-\omega_2^2\lambda^2A_Z-\frac{3}{2}C_3(-A_X(2k_{40}+k_{41})-k_{32}(\rho_{21}
	-\rho_{22})-\rho_{31}k_{21}-A_Z(2\rho_{40}+\rho_{41}))-\frac{3}{2}C_4(-A_X^2
	 k_{32}+4A_X(\rho_{20}(k_{21}+2k_{22})+(\rho_{21}-\rho_{22})(k_{21}
	+k_{22}))+2A_XA_Z\rho_{31}-2A_Z\{2\rho_{20}(\rho_{20}+\rho_{21}
	-\rho_{22})+ (\rho_{21} - \rho_{22})^2\}-\frac{1}{4}k^2A_X^2k_{32}+kA_X
	k_{22}(\sigma_{21}-\sigma_{22})
	+\frac{1}{2}kA_XA_Z(\sigma_{31}+\sigma_{32})
	+\frac{3}{4}A_Z^2k_{32}+\frac{1}{2}A_Z(\sigma_{21}
	-\sigma_{22})^2+\frac{3}{2}A_Z(k_{21}^2
	+2k_{21}k_{22}+2k_{22}^2))-\frac{5}{8}C_5(4A_X^3 (2k_{21}
	+3k_{22})-12A_X^2A_Z(3\rho_{20}+2(\rho_{21}-\rho_{22}))
	-9A_XA_Z^2(2k_{21}+3k_{22})+3A_Z^3(3\rho_{20}+2(\rho_{21}
	-\rho_{22}))         -3k^2A_X^3k_{22}-6kA_X^2A_Z(\sigma_{21}-\sigma_{22})
	+3k^2A_X^2A_Z\rho_{20})+\frac{15}{64}C_6(40A_X^4A_Z
	-12k^2A_X^4A_Z
	-60A_X^2A_Z^3+k^4A_X^4A_Z
	+2k^2A_X^2A_Z^3+5A_Z^5).
\end{dmath*}
\begin{align*}
\gamma_8 = \left\lbrace 
\begin{array}{c}
\gamma_{81}, \quad\text{when}\,\,p=0,2,\\
\gamma_{82}, \quad\text{when}\,\,p=1,3.
\end{array}
\right. 
\end{align*}		
\begin{dmath*}
	\gamma_{81} =
	6\lambda\omega_2(3\lambda\rho_{31}+n\sigma_{31})
	+\frac{3}{2}C_3(-2A_X(\rho_{41}+\rho_{42})+4\rho_{20}\rho_{31}+kA_X(\sigma_{41}
	-\sigma_{42})+\sigma_{32}(\sigma_{21}+\sigma_{22})+(-1)^{\frac{p}{2}}A_Z(k_{41}-k_{42}))
	+\frac{3}{2}C_4(4A_X^2\rho_{31}-2A_X(\rho_{21}+\rho_{22})(4\rho_{20}+\rho_{21}+\rho_{22})-2kA_X^2\sigma_{32}-A_X(\sigma_{21}
	+\sigma_{22})^2+2kA_X(\sigma_{21}+\sigma_{22})(2\rho_{20}-(\rho_{21}+\rho_{22}))-2k^2A_X^2\rho_{31}-A_Xk_{21}^2
	+2A_Zk_{21}(2\rho_{20}-(\rho_{21}+\rho_{22}))-2A_Z^2\rho_{31})+\frac{5}{16}C_5(-8A_X^3(2\rho_{20}
	+3(\rho_{21}+\rho_{22}))+12kA_X^3(\sigma_{21}+\sigma_{22})-12k^2A_X^3(2\rho_{20}-(\rho_{21}+\rho_{22}))
	+12A_X^2A_Zk_{21}-12A_XA_Z^2(2\rho_{20}-(\rho_{21}+\rho_{22}))-9k^3A_X^3(\sigma_{21}+\sigma_{22})-9k^2A_X^2A_Zk_{21}
	-9kA_XA_Z^2(\sigma_{21}+\sigma_{22})-9A_Z^3k_{21})+\frac{15}{64}C_6(-8A_X^5-8k^2A_X^5-8A_X^3A_Z^2+9k^4A_X^5
	+18k^2A_X^3A_Z^2+9A_XA_Z^4),
\end{dmath*}
\begin{dmath*}
	\gamma_{82} =
	6\lambda\omega_2(3\lambda\rho_{31}+n\sigma_{31})
	+\frac{3}{2}C_3(-2A_X(\rho_{41}+\rho_{42})+4\rho_{20}\rho_{31}+kA_X(\sigma_{41}-\sigma_{42})
	+\sigma_{32}(\sigma_{21}-\sigma_{22})-(-1)^{\frac{p-1}{2}}A_Z(k_{41}+k_{42})-2k_{22}k_{32})
	+\frac{3}{2}C_4(4A_X^2\rho_{31}-2A_X(\rho_{21}-\rho_{22})(4\rho_{20}+\rho_{21}-\rho_{22})
	-2kA_X^2\sigma_{32}-A_X(\sigma_{21}-\sigma_{22})^2+2kA_X(\sigma_{21}-\sigma_{22})(2\rho_{20}-(\rho_{21}-\rho_{22}))-2k^2A_X^2\rho_{31}
	+4A_XA_Zk_{32}+A_X(k_{21}^2+4k_{21}k_{22})-2A_Z(k_{21}(2\rho_{20}+\rho_{21}-\rho_{22})+2k_{22}(\rho_{21}-\rho_{22}))
	-2A_Z^2\rho_{31})+\frac{5}{16}C_5(-8A_X^3(2\rho_{20}+3(\rho_{21}-\rho_{22}))+12kA_X^3(\sigma_{21}-\sigma_{22})
	-12k^2A_X^3(2\rho_{20}-(\rho_{21}-\rho_{22}))-12A_X^2A_Z(3k_{21}+2k_{22})
	+12A_XA_Z^2(2\rho_{20}+3(\rho_{21}-\rho_{22}))-9k^3A_X^3(\sigma_{21}-\sigma_{22})+3k^2A_X^2A_Z(k_{21}-2k_{22})
	-3kA_XA_Z^2(\sigma_{21}-\sigma_{22})+3A_Z^3(3k_{21}+2k_{22}))
	+\frac{15}{64}C_6(-8A_X^5-8k^2A_X^5+40A_X^3A_Z^2+9k^4A_X^5
	+6k^2A_X^3A_Z^2-15A_XA_Z^4).
\end{dmath*}
\begin{align*}
\gamma_9 = \left\lbrace 
\begin{array}{c}
\gamma_{91}, \quad\text{when}\,\,p=0,2,\\
\gamma_{92}, \quad\text{when}\,\,p=1,3.
\end{array}
\right. 
\end{align*}		
\begin{dmath*}
	\gamma_{91} =\frac{3}{2}C_3(-2A_X\rho_{42}+2\rho_{31}(\rho_{21}+\rho_{22})+kA_X\sigma_{42}+\sigma_{31}(\sigma_{21}+\sigma_{22})
	+(-1)^{\frac{p}{2}}A_Zk_{42}+k_{21}k_{31})+\frac{3}{2}C_4(2A_X^2\rho_{31}-2A_X(\rho_{21}+\rho_{22})^2-2kA_X^2\sigma_{31}
	-A_X(\sigma_{21}+\sigma_{22})^2+2kA_X(\sigma_{21}+\sigma_{22})(\rho_{21}+\rho_{22})+k^2A_X^2\rho_{31}-2A_XA_Zk_{31}
	-A_Xk_{21}^2+2A_Zk_{21}(\rho_{21}+\rho_{22})+A_Z^2\rho_{31})
	+\frac{5}{16}C_5(-8A_X^3(\rho_{21}+\rho_{22})+12kA_X^3(\sigma_{21}+\sigma_{22})-12k^2A_X^3(\rho_{21}+\rho_{22})
	+12A_X^2A_Zk_{21}-12A_XA_Z^2(\rho_{21}+\rho_{22})+3k^3A_X^3(\sigma_{21}+\sigma_{22})+3k^2A_X^2A_Zk_{21}
	+3kA_XA_Z^2(\sigma_{21}+\sigma_{22})+3A_Z^2k_{21})+\frac{3}{64}C_6(-8A_X^5-40k^2A_X^5-40A_X^3A_Z^2-15k^4A_X^5
	-30k^2A_X^3A_Z^2-15A_XA_Z^4),
\end{dmath*}
\begin{dmath*}
	\gamma_{92} =\frac{3}{2}C_3(-2A_X\rho_{42}+2\rho_{31}(\rho_{21}-\rho_{22})+kA_X\sigma_{42}+\sigma_{31}(\sigma_{21}-\sigma_{22})
	-(-1)^{\frac{p-1}{2}}A_Zk_{42}-k_{21}k_{31})+\frac{3}{2}C_4(2A_X^2\rho_{31}-2A_X(\rho_{21}-\rho_{22})^2-2kA_X^2\sigma_{31}
	-A_X(\sigma_{21}-\sigma_{22})^2+2kA_X(\sigma_{21}-\sigma_{22})(\rho_{21}-\rho_{22})+k^2A_X^2\rho_{31}+2A_XA_Zk_{32}
	+A_Xk_{21}^2-2A_Zk_{21}(\rho_{21}-\rho_{22})-A_Z^2\rho_{31})
	+\frac{5}{16}C_5(-8A_X^3(\rho_{21}-\rho_{22})+12kA_X^3(\sigma_{21}-\sigma_{22})-12k^2A_X^3(\rho_{21}-\rho_{22})
	-12A_X^2A_Zk_{21}+12A_XA_Z^2(\rho_{21}-\rho_{22})+3k^3A_X^3(\sigma_{21}-\sigma_{22})-3k^2A_X^2A_Zk_{21}
	-3kA_XA_Z^2(\sigma_{21}-\sigma_{22})+3A_Z^2k_{21})+\frac{3}{64}C_6(-8A_X^5-40k^2A_X^5+40A_X^3A_Z^2-15k^4A_X^5
	+30k^2A_X^3A_Z^2-15A_XA_Z^4).
\end{dmath*}
\begin{align*}
\beta_9 = \left\lbrace 
\begin{array}{c}
\beta_{91}, \quad\text{when}\,\,p=0,2,\\
\beta_{92}, \quad\text{when}\,\,p=1,3.
\end{array}
\right. 
\end{align*}		
\begin{dmath*}
	\beta_{91} = 6\lambda\omega_2(3\lambda\sigma_{31}+n\rho_{31})
	-\frac{3}{2}C_3(-A_X(\sigma_{41}+\sigma_{42})+2\rho_{20}\sigma_{31}+\sigma_{32}(\rho_{21}+\rho_{22})+kA_X(\rho_{41}-\rho_{42}))
	-\frac{3}{8}C_4(4A_X^2(2\sigma_{31}+\sigma_{32})-8A_X(\sigma_{21}+\sigma_{22})(2\rho_{20}+\rho_{21}+\rho_{22})
	+4kA_X(\rho_{21}+\rho_{22})(4\rho_{20}-(\rho_{21}+\rho_{22}))-3k^2A_X^2(2\sigma_{31}-\sigma_{32})-3kA_X(\sigma_{21}+\sigma_{22})^2
	-4kA_XA_Zk_{31}-kA_Xk_{21}^2-2A_Zk_{21}(\sigma_{21}+\sigma_{22})
	-A_Z^2(2\sigma_{31}-\sigma_{32}))
	+\frac{5}{16}C_5(12A_X^3(\sigma_{21}+\sigma_{22})-12kA_X^3(2\rho_{20}+\rho_{21}+\rho_{22})+3k^3A_X^3(-2\rho_{20}+3(\rho_{21}+\rho_{22}))
	-9k^2A_X^3(\sigma_{21}+\sigma_{22})-6kA_X^2A_Zk_{21}-3A_XA_Z^2(\sigma_{21}+\sigma_{22})
	+3kA_XA_Z^2(-2\rho_{20}+\rho_{21}+\rho_{22}))+\frac{15}{128}C_6(-24kA_X^5+12k^3A_X^5
	+12kA_X^3A_Z^2+5k^5A_X^5+10k^3A_X^3A_Z^2+5kA_XA_Z^4), 
\end{dmath*}
\begin{dmath*}
	\beta_{92} = 6\lambda\omega_2(3\lambda\sigma_{31}+n\rho_{31})
	-\frac{3}{2}C_3(-A_X(\sigma_{41}+\sigma_{42})+2\rho_{20}\sigma_{31}+\sigma_{32}(\rho_{21}-\rho_{22})+kA_X(\rho_{41}-\rho_{42}))
	-\frac{3}{8}C_4(4A_X^2(2\sigma_{31}+\sigma_{32})-8A_X(\sigma_{21}-\sigma_{22})(2\rho_{20}+\rho_{21}-\rho_{22})
	+4kA_X(\rho_{21}-\rho_{22})(4\rho_{20}-(\rho_{21}-\rho_{22}))-3k^2A_X^2(2\sigma_{31}+\sigma_{32})-3kA_X(\sigma_{21}-\sigma_{22})^2
	-kA_X(4k_{21}k_{22}-k_{21}^2)-2A_Z(k_{21}+2k_{22})(\sigma_{21}-\sigma_{22})
	-A_Z^2(2\sigma_{31}+\sigma_{32}))
	+\frac{5}{16}C_5(12A_X^3(\sigma_{21}-\sigma_{22})-12kA_X^3(2\rho_{20}+\rho_{21}-\rho_{22})+3k^3A_X^3(-2\rho_{20}+3(\rho_{21}-\rho_{22}))
	-9k^2A_X^3(\sigma_{21}-\sigma_{22})-6kA_X^2A_Z(k_{21}+2k_{22})-9A_XA_Z^2(\sigma_{21}-\sigma_{22})
	+3kA_XA_Z^2(2\rho_{20}+\rho_{21}-\rho_{22}))+\frac{15}{128}C_6(-24kA_X^5+12k^3A_X^5\\
	+36kA_X^3A_Z^2+5k^5A_X^5-2k^3A_X^3A_Z^2-3kA_XA_Z^4).
\end{dmath*}
\begin{align*}
\beta_{10} = \left\lbrace 
\begin{array}{c}
\beta_{101}, \quad\text{when}\,\,p=0,2,\\
\beta_{102}, \quad\text{when}\,\,p=1,3.
\end{array}
\right. 
\end{align*}		
\begin{dmath*}
	\beta_{101} = 
	-\frac{3}{2}C_3(-A_X\sigma_{42}+\sigma_{31}(\rho_{21}+\rho_{22})+\rho_{31}(\sigma_{21}+\sigma_{22})+kA_X\rho_{42})
	-\frac{3}{8}C_4(4A_X^2\sigma_{31}-8A_X(\sigma_{21}+\sigma_{22})(\rho_{21}+\rho_{22})-8kA_X^2\rho_{31}+4kA_X(\rho_{21}+\rho_{22})^2
	+3k^2A_X^2\sigma_{31}
	+3kA_X(\sigma_{21}+\sigma_{22})^2+2kA_XA_Zk_{31}+kA_Xk_{21}^2+2A_Zk_{21}(\sigma_{21}+\sigma_{22})
	+A_Z^2\sigma_{31})
	+\frac{5}{16}C_5(4A_X^3(\sigma_{21}+\sigma_{22})-12kA_X^3(\rho_{21}+\rho_{22})-3k^3A_X^3(\rho_{21}+\rho_{22})
	+9k^2A_X^3(\sigma_{21}+\sigma_{22})+6kA_X^2A_Zk_{21}+3A_XA_Z^2(\sigma_{21}+\sigma_{22})-3kA_XA_Z^2(\rho_{21}+\rho_{22}))
	+\frac{15}{128}C_6(-8kA_X^5-12k^3A_X^5-12kA_X^3A_Z^2-k^5A_X^5
	-2k^3A_X^3A_Z^2-kA_XA_Z^4),
\end{dmath*}
\begin{dmath*}
	\beta_{102} = -\frac{3}{2}C_3(-A_X\sigma_{42}+\sigma_{31}(\rho_{21}-\rho_{22})+\rho_{31}(\sigma_{21}-\sigma_{22})+kA_X\rho_{42})
	-\frac{3}{8}C_4(4A_X^2\sigma_{31}-8A_X(\sigma_{21}-\sigma_{22})(\rho_{21}-\rho_{22})-8kA_X^2\rho_{31}+4kA_X(\rho_{21}-\rho_{22})^2
	+3k^2A_X^2\sigma_{31}
	+3kA_X(\sigma_{21}-\sigma_{22})^2-2kA_XA_Zk_{32}-kA_Xk_{21}^2-2A_Zk_{21}(\sigma_{21}-\sigma_{22})
	-A_Z^2\sigma_{31})
	+\frac{5}{16}C_5(4A_X^3(\sigma_{21}-\sigma_{22})-12kA_X^3(\rho_{21}-\rho_{22})-3k^3A_X^3(\rho_{21}-\rho_{22})
	+9k^2A_X^3(\sigma_{21}-\sigma_{22})-6kA_X^2A_Zk_{21}-3A_XA_Z^2(\sigma_{21}-\sigma_{22})+3kA_XA_Z^2(\rho_{21}-\rho_{22}))
	+\frac{15}{128}C_6(-8kA_X^5-12k^3A_X^5+12kA_X^3A_Z^2-k^5A_X^5
	-2k^3A_X^3A_Z^2-kA_XA_Z^4).
\end{dmath*}
\begin{align*}
\delta_8 = \left\lbrace 
\begin{array}{c}
\delta_{80}, \quad\text{when}\,\,p=0,\\
\delta_{81}, \quad\text{when}\,\,p=1,\\
\delta_{82},\quad\text{when}\,\,p=2,\\
\delta_{83},\quad\text{when} \,\,p=3.
\end{array}
\right. 
\end{align*}		
\begin{dmath*}
	\delta_{80}=18\omega_2\lambda^2k_{31}-\frac{3}{2}C_3(-A_X(k_{41}+k_{42})+A_Z(\rho_{41}-\rho_{42})
	+2\rho_{20}k_{31})-\frac{3}{8}C_4(8A_X^2k_{31}-8A_Xk_{21}(2\rho_{20}+\rho_{21}+\rho_{22})\\
	+4A_Z(\rho_{21}+\rho_{22})(4\rho_{20}-\rho_{21}-\rho_{22})
	-2k^2A_X^2k_{31}-2kA_Xk_{21}(\sigma_{21}+\sigma_{22})-2kA_XA_Z(2\sigma_{31}-\sigma_{32})-6A_Z^2k_{31}
	-A_Z(\sigma_{21}+\sigma_{22})^2
	-3A_Zk_{21}^2)-\frac{5}{16}C_5(-12A_X^3k_{21}+12A_X^2A_Z(2\rho_{20}+\rho_{21}+\rho_{22})
	+9A_XA_Z^2k_{21}+3A_Z^3(2\rho_{20}
	-3(\rho_{21}+\rho_{22}))+3k^2A_X^3k_{21}-6kA_X^2A_Z(\sigma_{21}+\sigma_{22})
	+3k^2A_X^2A_Z(2\rho_{20}-3(\rho_{21}+\rho_{22})))
	+\frac{15}{128}C_6(-24A_X^4A_Z+12k^2A_X^4A_Z
	+12A_X^2A_Z^3+5k^4A_X^4A_Z+10k^2A_X^2A_Z+5A_Z^5)+\frac{\Delta}{\epsilon^2}k_{31},
\end{dmath*}
\begin{dmath*}
	\delta_{81}=18\omega_2\lambda^2k_{32}-\frac{3}{2}C_3(-A_X(k_{41}+k_{42})+A_Z(\rho_{41}+\rho_{42})+2\rho_{31}k_{22}
	+2\rho_{20}k_{32})-\frac{3}{8}C_4(8A_X^2k_{32}-8A_X(2\rho_{20}k_{21}+(\rho_{21}-\rho_{22})(k_{21}+2k_{22}))
	-16A_XA_Z\rho_{31}+4A_Z(\rho_{21}-\rho_{22})(4\rho_{20}+\rho_{21}-\rho_{22})\\
	-2k^2A_X^2k_{32}-2kA_X(k_{21}-2k_{22})(\sigma_{21}-\sigma_{22})+2kA_XA_Z\sigma_{32}-6A_Z^2k_{32}
	+A_Z(\sigma_{21}-\sigma_{22})^2
	-3A_Zk_{21}(k_{21}+4k_{22}))-\frac{5}{16}C_5(-4A_X^3(3k_{21}+2k_{22})\\+12A_X^2A_Z(2\rho_{20}+3(\rho_{21}-\rho_{22}))
	+9A_XA_Z^2(3k_{21}+2k_{22})-3A_Z^3(2\rho_{20}
	+3(\rho_{21}-\rho_{22}))
	+3k^2A_X^3(k_{21}-2k_{22})-6kA_X^2A_Z(\sigma_{21}-\sigma_{22})
	+3k^2A_X^2A_Z(2\rho_{20}-(\rho_{21}-\rho_{22})))
	+\frac{15}{128}C_6(-40A_X^4A_Z-12k^2A_X^4A_Z
	+60A_X^2A_Z^3+3k^4A_X^4A_Z+2k^2A_X^2A_Z^3-A_Z^5)+\frac{\Delta}{\epsilon^2}k_{32},
\end{dmath*}
\begin{dmath*}
	\delta_{82}=-18\omega_2\lambda^2k_{31}-\frac{3}{2}C_3(-A_X(k_{41}+k_{42})-A_Z(\rho_{41}-\rho_{42})
	-2\rho_{20}k_{31})-\frac{3}{8}C_4(-8A_X^2k_{31}+8A_Xk_{21}(2\rho_{20}+\rho_{21}+\rho_{22})
	-4A_Z(\rho_{21}+\rho_{22})(4\rho_{20}-\rho_{21}-\rho_{22})
	+2k^2A_X^2k_{31}+2kA_Xk_{21}(\sigma_{21}+\sigma_{22})+2kA_XA_Z(2\sigma_{31}-\sigma_{32})+6A_Z^2k_{31}
	+A_Z(\sigma_{21}+\sigma_{22})^2
	+3A_Zk_{21}^2)-\frac{5}{16}C_5(12A_X^3k_{21}-12A_X^2A_Z(2\rho_{20}+\rho_{21}+\rho_{22})
	-9A_XA_Z^2k_{21}-3A_Z^3(2\rho_{20}
	-3(\rho_{21}+\rho_{22}))-3k^2A_X^3k_{21}+6kA_X^2A_Z(\sigma_{21}+\sigma_{22})
	-3k^2A_X^2A_Z(2\rho_{20}-3(\rho_{21}+\rho_{22})))
	+\frac{15}{128}C_6(24A_X^4A_Z-12k^2A_X^4A_Z
	-12A_X^2A_Z^3-5k^4A_X^4A_Z-10k^2A_X^2A_Z^3-5A_Z^5)-\frac{\Delta}{\epsilon^2}k_{31},
\end{dmath*}
\begin{dmath*}
	\delta_{83}=-18\omega_2\lambda^2k_{32}-\frac{3}{2}C_3(-A_X(k_{41}+k_{42})-A_Z(\rho_{41}+\rho_{42})-2\rho_{31}k_{22}
	-2\rho_{20}k_{32})-\frac{3}{8}C_4(-8A_X^2k_{32}+8A_X(2\rho_{20}k_{21}+(\rho_{21}-\rho_{22})(k_{21}+2k_{22}))
	+16A_XA_Z\rho_{31}-4A_Z(\rho_{21}-\rho_{22})(4\rho_{20}+\rho_{21}-\rho_{22})
	+2k^2A_X^2k_{32}+2kA_X(k_{21}-2k_{22})(\sigma_{21}-\sigma_{22})-2kA_XA_Z\sigma_{32}+6A_Z^2k_{32}
	-A_Z(\sigma_{21}-\sigma_{22})^2
	+3A_Zk_{21}(k_{21}+4k_{22}))-\frac{5}{16}C_5(4A_X^3(3k_{21}+2k_{22})\\-12A_X^2A_Z(2\rho_{20}+3(\rho_{21}-\rho_{22}))
	-9A_XA_Z^2(3k_{21}+2k_{22})+3A_Z^3(2\rho_{20}
	+3(\rho_{21}-\rho_{22}))
	-3k^2A_X^3(k_{21}-2k_{22})+6kA_X^2A_Z(\sigma_{21}-\sigma_{22})
	-3k^2A_X^2A_Z(2\rho_{20}-(\rho_{21}-\rho_{22})))
	+\frac{15}{128}C_6(40A_X^4A_Z+12k^2A_X^4A_Z
	-60A_X^2A_Z^3-3k^4A_X^4A_Z-2k^2A_X^2A_Z^3+5A_Z^5)-\frac{\Delta}{\epsilon^2}k_{32}.
\end{dmath*}
\begin{align*}
\delta_9 = \left\lbrace 
\begin{array}{c}
\delta_{90}, \quad\text{when}\,\,p=0,\\
\delta_{91}, \quad\text{when}\,\,p=1,\\
\delta_{92},\quad\text{when}\,\,p=2,\\
\delta_{93},\quad\text{when} \,\,p=3.
\end{array}
\right. 
\end{align*}		
\begin{dmath*}
	\delta_{90}=-\frac{3}{2}C_3(-A_Xk_{42}+k_{31}(\rho_{21}+\rho_{22})+\rho_{31}k_{21}
	+A_Z\rho_{42})-\frac{3}{8}C_4(4A_X^2k_{31}-8A_Xk_{21}(\rho_{21}
	+\rho_{22})-8A_XA_Z\rho_{31}+4A_Z(\rho_{21}+\rho_{22})^2+k^2A_X^2k_{31}
	+2kA_Xk_{21}(\sigma_{21}+\sigma_{22})+2kA_XA_Z\sigma_{31}+3A_Z^2k_{31}
	+A_Z(\sigma_{21}+\sigma_{22})^2+3A_Zk_{21}^2)-\frac{5}{16}C_5(-4A_X^3k_{21}
	+12A_X^2A_Z(\rho_{21}+\rho_{22})-9A_XA_Z^2k_{21}+3A_Z^3(\rho_{21}+\rho_{22})
	-3k^2A_X^3k_{21}+6kA_X^2A_Z(\sigma_{21}+\sigma_{22})+3k^2A_X^2A_Z(\rho_{21}
	+\rho_{22}))+\frac{15}{128}C_6(-8A_X^4A_Z-12k^2A_X^4A_Z
	-12A_X^2A_Z^3-k^4A_X^4A_Z
	-2k^2A_X^2A_Z^3-A_Z^5),
\end{dmath*}
\begin{dmath*}
	\delta_{91}=-\frac{3}{2}C_3(-A_Xk_{42}+k_{32}(\rho_{21}-\rho_{22})+\rho_{31}k_{21}
	+A_Z\rho_{42})-\frac{3}{8}C_4(4A_X^2k_{32}-8A_Xk_{21}(\rho_{21}-\rho_{22})
	-8A_XA_Z\rho_{31}+4A_Z(\rho_{21}-\rho_{22})^2+k^2A_X^2k_{32}+2kA_Xk_{21}(\sigma_{21}-\sigma_{22})
	+2kA_XA_Z\sigma_{31}-3A_Z^2k_{32}+A_Z(\sigma_{21}-\sigma_{22})^2-3A_Zk_{21}^2)
	-\frac{5}{16}C_5(-4A_X^3k_{21}+12A_X^2A_Z(\rho_{21}-\rho_{22})+9A_XA_Z^2k_{21}-3A_Z^3(\rho_{21}-\rho_{22})
	-3k^2A_X^3k_{21}-6kA_X^2A_Z(\sigma_{21}-\sigma_{22})+3k^2A_X^2A_Z(\rho_{21}
	-\rho_{22}))+\frac{15}{128}C_6(-8A_X^4A_Z-12k^2A_X^4A_Z
	+12A_X^2A_Z^3-k^4A_X^4A_Z+2k^2A_X^2A_Z^3-A_Z^5),
\end{dmath*}
\begin{dmath*}
	\delta_{92}=-\frac{3}{2}C_3(-A_Xk_{42}-k_{31}(\rho_{21}+\rho_{22})-\rho_{31}k_{21}
	-A_Z\rho_{42})-\frac{3}{8}C_4(-4A_X^2k_{31}+8A_Xk_{21}(\rho_{21}+\rho_{22})
	+8A_XA_Z\rho_{31}-4A_Z(\rho_{21}+\rho_{22})^2-k^2A_X^2k_{31}-2kA_Xk_{21}(\sigma_{21}+\sigma_{22})
	-2kA_XA_Z\sigma_{31}-3A_Z^2k_{31}-A_Z(\sigma_{21}+\sigma_{22})^2-3A_Zk_{21}^2)
	-\frac{5}{16}C_5(4A_X^3k_{21}-12A_X^2A_Z(\rho_{21}+\rho_{22})+9A_XA_Z^2k_{21}-3A_Z^3(\rho_{21}+\rho_{22})
	+3k^2A_X^3k_{21}-6kA_X^2A_Z(\sigma_{21}+\sigma_{22})-3k^2A_X^2A_Z(\rho_{21}
	+\rho_{22}))+\frac{15}{128}C_6(8A_X^4A_Z+12k^2A_X^4A_Z
	+12A_X^2A_Z^3+k^4A_X^4A_Z+2k^2A_X^2A_Z^3+A_Z^5),
\end{dmath*}
\begin{dmath*}
	\delta_{93}=-\frac{3}{2}C_3(-A_Xk_{42}-k_{32}(\rho_{21}-\rho_{22})-\rho_{31}k_{21}
	-A_Z\rho_{42})-\frac{3}{8}C_4(-4A_X^2k_{32}+8A_Xk_{21}(\rho_{21}-\rho_{22})
	+8A_XA_Z\rho_{31}-4A_Z(\rho_{21}-\rho_{22})^2-k^2A_X^2k_{32}-2kA_Xk_{21}(\sigma_{21}-\sigma_{22})
	-2kA_XA_Z\sigma_{31}+3A_Z^2k_{32}-A_Z(\sigma_{21}-\sigma_{22})^2+3A_Zk_{21}^2)
	-\frac{5}{16}C_5(4A_X^3k_{21}-12A_X^2A_Z(\rho_{21}-\rho_{22})-9A_XA_Z^2k_{21}+3A_Z^3(\rho_{21}-\rho_{22})
	+3k^2A_X^3k_{21}+6kA_X^2A_Z(\sigma_{21}-\sigma_{22})-3k^2A_X^2A_Z(\rho_{21}
	-\rho_{22}))+\frac{15}{128}C_6(8A_X^4A_Z+12k^2A_X^4A_Z
	-12A_X^2A_Z^3+k^4A_X^4A_Z-2k^2A_X^2A_Z^3+A_Z^5).
\end{dmath*}
\begin{align*}
\rho_{51} =& \frac{6n\lambda\beta_9-(9\lambda^2+n^2-C_2)\gamma_8}{(n^2-9\lambda^2)^2+C_2(n^2-2C_2+9\lambda^2)}_,\\
\rho_{52} =& \frac{10n\lambda\beta_{10}-(25\lambda^2+n^2-C_2)\gamma_9}{(n^2-25\lambda^2)^2+C_2(n^2-2C_2+25\lambda^2)}_,\\ 
\sigma_{51} = & -\frac{k\beta_{11}}{2\lambda n}_,\\
\sigma_{52} =& \frac{6n\lambda\gamma_8-(9\lambda^2+n^2+2C_2)\beta_9}{(n^2-9\lambda^2)^2+C_2(n^2-2C_2+9\lambda^2)}_,\\ 
\sigma_{53} =& \frac{10n\lambda\gamma_9-(25\lambda^2+n^2+2C_2)\beta_{10}}{(n^2-25\lambda^2)^2+C_2(n^2-2C_2+25\lambda^2)}_,\\
k_{51} =& -\frac{\delta_8}{8\lambda^2}_,\\
k_{52} =& -\frac{\delta_9}{24\lambda^2}_.
\end{align*}

\begin{table*}[h!]
	\footnotesize
	\centering
	\caption{Effect of radiation pressure on different parameters of orbits around $L_1$ when $A_2 = 2.4\times 10^{-12}$} 
	\label{tbl:1}
	\begin{tabular}{|l||l|l|l|l|l|}
		\tableline  
		
		$q$& 1 & 0.9995 & 0.9945 & 0.9895 & 0.9845 \\ \tableline
		$\gamma$& 0.009966562831474 &  0.010022806042997 & 0.010621857046575 & 0.011292634839768 & 0.012042049799192 \\ \tableline
	
		$L_1$& 0.990030433658191 & 0.989974190565718 & 0.989375139443090 & 0.988704361649897 & 0.987954946690473 \\ \tableline
		
		$C_2$& 4.064344359358607 & 4.013217307075904 & 3.533142166508352 & 3.109449638463373  & 2.740934989090651 \\ \tableline
		
		$C_3$& 3.023467759599181 & 2.972624703974638 & 2.495245216485259 & 2.073965581138543 & 1.707553745778144 \\ \tableline
		
		$C_4$& 3.033946155218810 & 2.983159954578718 & 2.506387966212186 & 2.085792481345639 & 1.720149452296818 \\ \tableline
		
		$C_5$& 3.033840670308508 & 2.983053292754523 & 2.506268338852770 & 2.085657399045222 & 1.719995925393043 \\ \tableline
		
		$C_6$& 3.033841732214042 & 2.983054372628669 & 2.506269623159199 & 2.085658941903247 & 1.719997796706109 \\ \tableline
		
		$\lambda$& 2.087246092417118 & 2.074819527399089 & 1.954036850904442 & 1.840326271553855 & 1.734687871104357\\ \tableline
		
		$\Delta$& 0.292251890951922 & 0.291658764200673 & 0.285117848184197 & 0.277351147307939 & 0.268207021065916\\ \tableline
		 
		 $k$& 3.230401297192164 & 3.212643439431957 & 3.041023597428729 & 2.881472765622095 & 2.735654104240122 \\ \tableline
		 
		 $\tau$& 3.010275276119164 & 3.028304497912615 & 3.215489669128482 & 3.414169217871602 & 3.622084071631579\\ \tableline
		  	  
		\tableline 
	\end{tabular}
\end{table*}

\begin{table*}[h!]
	\footnotesize
	\centering
	\caption{Effect of oblateness on different parameters of orbits around $L_1$ when $q = 0.9995$} 
	\label{tbl:2}
	\begin{tabular}{|l||l|l|l|l|l|}
		\tableline  
		
		$A_2$ & 0 & $2.4\times 10^{-12}$ & $3\times 10^{-12}$ & $3.5\times 10^{-12}$ & $4\times 10^{-12}$ \\ \tableline
		$\gamma$ & 0.010022805923947 & 0.010022806042997 & 0.010022806072760 & 0.010022806097562 &0.010022806122364 \\ \tableline
		
		$L_1$ & 0.989974190565718 & 0.989974190446668 & 0.989974190416905 & 0.989974190392103 &0.989974190367301 \\ \tableline
		
		$C_2$ & 4.013217306099824 & 4.013217307075904 & 4.013217307319478  & 4.013217307522901 & 4.013217307726325 \\ \tableline
		
		$C_3$ & 2.972624703499105 & 2.972624703974638 & 2.972624704093073 & 2.972624704192215 & 2.972624704291359 \\ \tableline
		
		$C_4$ & 2.983159953971714 & 2.983159954578718 & 2.983159954730022 & 2.983159954856554 & 2.983159954983087 \\ \tableline
		
		$C_5$ &2.983053292150130 & 2.983053292754523 & 2.983053292905174 & 2.983053293031162 &2.983053293157152 \\ \tableline
		
		$C_6$ & 2.983054372024236 & 2.983054372628669 & 2.983054372779329 & 2.983054372905326 & 2.983054373031323\\ \tableline
		
		$\lambda$ &2.074819527160888 & 2.074819527399089 & 2.074819527458530 & 2.074819527508173 &2.074819527557816 \\ \tableline
		
		$\Delta$ & 0.291658764188306 & 0.291658764200673 & 0.291658764203759 & 0.291658764206337 &0.291658764208914 \\ \tableline
		
		$k$ & 3.212643439097060 & 3.212643439431957 & 3.212643439515526 & 3.212643439585321 &3.212643439655118 \\ \tableline
		
		$\tau$ & 3.028304498260281 & 3.028304497912615 & 3.028304497825857 & 3.028304497753401 & 3.028304497680945\\ \tableline
		
		\tableline 
	\end{tabular}
\end{table*}

\begin{table*}[h!]
	\footnotesize
	\centering
	\caption{Effect of radiation  pressure on different parameters of orbits around $L_2$ when $A_2 = 2.4\times 10^{-12}$} 
	\label{tbl:3}
	\begin{tabular}{|l||l|l|l|l|l|}
		\tableline  
		
		$q$ & 1 & 0.9995 & 0.9945 & 0.9895 & 0.9845 \\ \tableline
		$\gamma$ & 0.010033228531910 & 0.009978343639533 & 0.009461048204950 & 0.008996749353809 & 0.008579422887427 \\ \tableline
		
		$L_2$ & 1.010030225021575 & 1.009975340129198
		
		 & 1.009458044694615 & 1.008993745843474 & 1.008576419377092 \\ \tableline
		
		$C_2$& 3.944259093871529 & 3.993273144291635 & 4.513389113295589 & 5.087772984797185  & 5.715729697751704 \\ \tableline
		
		$C_3$ & -2.983408088154385 & -3.032693790758545 & -3.555654417985380 & -4.133098257601873 & -4.764305916840534 \\ \tableline
		
		$C_4$& 2.973863414322725 & 3.023203497285466
		
		 & 3.546678168593691 & 4.124585872165059 & 4.756212685136259 \\ \tableline
		
		$C_5$& -2.973768601705565 & -3.023109735463630 & -3.546594039811956 & -4.124509971228582 & -4.756143840525967 \\ \tableline
		
		$C_6$& 2.973767659878476 & 3.023108809119334 & 3.546593251325406 & 4.124509294455637 & 4.756143254903247 \\ \tableline
		
		$\lambda$& 2.057933451632755 &2.069950687116124 & 2.193217663095480 & 2.321449398886311 & 2.453753890572465 \\ \tableline
		
		$\Delta$& 0.290830997477577 & 0.291422702800878 & 0.296814604418408 & 0.301354326792431 & 0.305178457747805 \\ \tableline
		
		$k$& 3.188540491593625 & 3.205690410467006 & 3.382469098698882 & 3.567743774450670 & 3.760028180060947 \\ \tableline
		
		$\tau$& 3.053152813175049 & 3.035427532785036 & 2.864825235043738 & 2.706578618596585 & 2.560642015207853 \\ \tableline
		
		\tableline 
	\end{tabular}
\end{table*}
 
\begin{table*}[h!]
	\footnotesize
	\centering
	\caption{Effect of oblateness on different parameters of orbits around $L_2$ when $q = 0.9995$} 
	\label{tbl:4}
	\begin{tabular}{|l||l|l|l|l|l|}
		\tableline  
		
		$A_2$ & 0 & $2.4\times 10^{-12}$ & $3\times 10^{-12}$ & $3.5\times 10^{-12}$ & $4\times 10^{-12}$ \\ \tableline
		$\gamma$ &  0.009978343518616 & 0.009978343639533 & 0.009978343669762 & 0.009978343694953 &0.009978343720144\\ \tableline
		
		$L_2$ & 1.009975340008281 & 1.009975340129198 & 1.009975340159427 & 1.009975340184618 &1.009975340209809\\ \tableline
		
		$C_2$ & 3.993273145236837 & 3.993273144291635 & 3.993273144055562  & 3.993273143858681 &3.993273143661800 \\ \tableline
		
		$C_3$ & -3.032693791243736 & -3.032693790758545 & -3.032693790637474 & -3.032693790536429 &-3.032693790435384 \\ \tableline
		
		$C_4$ & 3.023203497879979 & 3.023203497285466 & 3.023203497137065 &3.023203497013244 & 3.023203496889423\\ \tableline
		
		$C_5$ & -3.023109736060348 & -3.023109735463630 & -3.023109735314677 & -3.023109735190398 &-3.023109735066117 \\ \tableline
		
		$C_6$ & 3.023108809716084 & 3.023108809119334 & 3.023108808970372 & 3.023108808846086 &3.023108808721798 \\ \tableline
		
		$\lambda$ &2.069950687346945 & 2.069950687116124 & 2.069950687058475 & 2.069950687010396 &2.069950686962317 \\ \tableline
		
		$\Delta$ & 0.291422702811252 & 0.291422702800878
		
		 & 0.291422702798289 & 0.291422702796127 &0.291422702793967 \\ \tableline
		
		$k$ & 3.205690410801889 & 3.205690410467006 & 3.205690410383364 & 3.205690410313609 &3.205690410243853 \\ \tableline
		
		$\tau$ & 3.035427532446555 & 3.035427532785036 & 3.035427532869575 & 3.035427532940079 & 3.035427533010583\\ \tableline
		
		\tableline 
	\end{tabular}
\end{table*} 
 
%
\bibliographystyle{spr-mp-nameyear-cnd}  
\bibliography{ms}  

\begin{thebibliography}{31}
\ifx \bisbn   \undefined \def \bisbn  #1{ISBN #1}\fi
\ifx \binits  \undefined \def \binits#1{#1} \fi
\ifx \bauthor  \undefined \def \bauthor#1{#1} \fi
\ifx \batitle  \undefined \def \batitle#1{#1} \fi
\ifx \bjtitle  \undefined \def \bjtitle#1{#1}\fi
\ifx \bvolume  \undefined \def \bvolume#1{\textbf{#1}}\fi
\ifx \byear  \undefined \def \byear#1{#1} \fi
\ifx \bissue  \undefined \def \bissue#1{#1} \fi
\ifx \bfpage  \undefined \def \bfpage#1{#1} \fi
\ifx \blpage  \undefined \def \blpage #1{#1} \fi
\ifx \burl  \undefined \def \burl#1{\textsf{#1}} \fi
\ifx \doiurl  \undefined \def \doiurl#1{\textsf{#1}} \fi
\ifx \betal  \undefined \def \betal{\textit{et al.}} \fi
\ifx \binstitute  \undefined \def \binstitute#1{#1} \fi
\ifx \binstitutionaled  \undefined \def \binstitutionaled#1{#1} \fi
\ifx \bctitle  \undefined \def \bctitle#1{#1} \fi
\ifx \beditor  \undefined \def \beditor#1{#1} \fi
\ifx \bpublisher  \undefined \def \bpublisher#1{#1} \fi
\ifx \bbtitle  \undefined \def \bbtitle#1{#1} \fi
\ifx \bedition  \undefined \def \bedition#1{#1} \fi
\ifx \bseriesno  \undefined \def \bseriesno#1{#1} \fi
\ifx \blocation  \undefined \def \blocation#1{#1} \fi
\ifx \bsertitle  \undefined \def \bsertitle#1{#1} \fi
\ifx \bsnm \undefined \def \bsnm#1{#1} \fi
\ifx \bsuffix \undefined \def \bsuffix#1{#1} \fi
\ifx \bparticle \undefined \def \bparticle#1{#1} \fi
\ifx \barticle \undefined \def \barticle#1{#1} \fi
\ifx \bconfdate \undefined \def \bconfdate #1{#1} \fi
\ifx \botherref \undefined \def \botherref #1{#1} \fi
\ifx \url \undefined \def \url#1{\textsf{#1}} \fi
\ifx \bchapter \undefined \def \bchapter#1{#1} \fi
\ifx \bbook \undefined \def \bbook#1{#1} \fi
\ifx \bcomment \undefined \def \bcomment#1{#1} \fi
\ifx \oauthor \undefined \def \oauthor#1{#1} \fi
\ifx \citeauthoryear \undefined \def \citeauthoryear#1{#1} \fi
\ifx \endbibitem  \undefined \def \endbibitem {}\fi
\ifx \bconflocation  \undefined \def \bconflocation#1{#1} \fi
\ifx \arxivurl  \undefined \def \arxivurl#1{\textsf{#1}} \fi

\bibitem[\protect\citeauthoryear{Abouelmagd}{2013}]{Abouelmagd2013}
\begin{barticle}
\bauthor{\bsnm{Abouelmagd}, \binits{E.I.}}:
\bjtitle{Astrophysics and Space Science}
\bvolume{346}(\bissue{1}),
\bfpage{51}
(\byear{2013}).
doi:\doiurl{10.1007/s10509-013-1439-9}
\end{barticle}
\endbibitem

\bibitem[\protect\citeauthoryear{Breakwell and Brown}{1979}]{Breakwell}
\begin{barticle}
\bauthor{\bsnm{Breakwell}, \binits{J.V.}},
\bauthor{\bsnm{Brown}, \binits{J.V.}}:
\bjtitle{Celestial Mechanics}
\bvolume{20},
\bfpage{389}
(\byear{1979})
\end{barticle}
\endbibitem

\bibitem[\protect\citeauthoryear{Brouwer and Clemence}{1961}]{Brouwer&Clemence}
\begin{botherref}
\oauthor{\bsnm{Brouwer}, \binits{D.}},
\oauthor{\bsnm{Clemence}, \binits{G.M.}}:
Planets and Satellites
\textbf{31}
(1961)
\end{botherref}
\endbibitem

\bibitem[\protect\citeauthoryear{Chidambararaj and
  Sharma}{2016}]{Prithiviraj&Sharma2016}
\begin{barticle}
\bauthor{\bsnm{Chidambararaj}, \binits{P.}},
\bauthor{\bsnm{Sharma}, \binits{R.K.}}:
\bjtitle{International Journal of Astronomy and Astrophysics}
\bvolume{6},
\bfpage{293}
(\byear{2016}).
doi:\doiurl{10.4236/ijaa.2016.63025}
\end{barticle}
\endbibitem

\bibitem[\protect\citeauthoryear{Danby}{1964}]{Danby}
\begin{bbook}
\bauthor{\bsnm{Danby}, \binits{J.M.A.}}:
\bbtitle{Fundamentals of Celestial Mechanics}.
\bpublisher{Macmillan Company},
\blocation{New York}
(\byear{1964})
\end{bbook}
\endbibitem

\bibitem[\protect\citeauthoryear{Eapen and Sharma}{2014}]{Eapen2014}
\begin{barticle}
\bauthor{\bsnm{Eapen}, \binits{R.T.}},
\bauthor{\bsnm{Sharma}, \binits{R.K.}}:
\bjtitle{Astrophysics and Space Science}
\bvolume{352}(\bissue{2}),
\bfpage{437}
(\byear{2014}).
doi:\doiurl{10.1007/s10509-014-1951-6}
\end{barticle}
\endbibitem

\bibitem[\protect\citeauthoryear{Farquhar}{1968}]{Farquhar1968}
\begin{botherref}
\oauthor{\bsnm{Farquhar}, \binits{R.W.}}:
The control and use of libration-point satellites.
PhD thesis,
Department of Aeronautics and Astronautics, Stanford University, Stanford
(1968)
\end{botherref}
\endbibitem

\bibitem[\protect\citeauthoryear{Fitzpatrick}{2012}]{Fitzpatrick}
\begin{bbook}
\bauthor{\bsnm{Fitzpatrick}, \binits{R.}}:
\bbtitle{An Introduction to Celestial Mechanics}.
\bpublisher{Cambridge University Press},
\blocation{New York}
(\byear{2012})
\end{bbook}
\endbibitem

\bibitem[\protect\citeauthoryear{Ghotekar and Sharma}{2019}]{Saurav&Sharma2019}
\begin{barticle}
\bauthor{\bsnm{Ghotekar}, \binits{S.}},
\bauthor{\bsnm{Sharma}, \binits{R.K.}}:
\bjtitle{International Journal of Astronomy and Astrophysics}
\bvolume{9},
\bfpage{274}
(\byear{2019}).
doi:\doiurl{10.4236/ijaa.2019.93020}
\end{barticle}
\endbibitem

\bibitem[\protect\citeauthoryear{Howell}{1984}]{Howell1984}
\begin{barticle}
\bauthor{\bsnm{Howell}, \binits{K.C.}}:
\bjtitle{Celestial Mechanics}
\bvolume{32},
\bfpage{53}
(\byear{1984})
\end{barticle}
\endbibitem

\bibitem[\protect\citeauthoryear{Howell and V.Breakwell}{1984}]{Howell}
\begin{barticle}
\bauthor{\bsnm{Howell}, \binits{K.C.}},
\bauthor{\bsnm{V.Breakwell}, \binits{J.}}:
\bjtitle{Celestial Mechanics}
\bvolume{32},
\bfpage{29}
(\byear{1984})
\end{barticle}
\endbibitem

\bibitem[\protect\citeauthoryear{Koon et~al.}{2011}]{Koon}
\begin{bbook}
\bauthor{\bsnm{Koon}, \binits{W.S.}},
\bauthor{\bsnm{W.Lo}, \binits{M.}},
\bauthor{\bsnm{Marsden}, \binits{J.E.}},
\bauthor{\bsnm{Ross}, \binits{S.D.}}:
\bbtitle{Dynamical Systems, the Three-body Problem and Space Mission Design}.
\bpublisher{Interdisplinary Applied Mathemstics, Springer},
\blocation{Berlin}
(\byear{2011})
\end{bbook}
\endbibitem

\bibitem[\protect\citeauthoryear{McCuskey}{1963}]{McCuskey}
\begin{bbook}
\bauthor{\bsnm{McCuskey}, \binits{S.W.}}:
\bbtitle{Introduction to Celestial Mechanics}.
\bpublisher{Addison-Wesley},
\blocation{London}
(\byear{1963})
\end{bbook}
\endbibitem

\bibitem[\protect\citeauthoryear{Moulton}{1914}]{Moulton}
\begin{bbook}
\bauthor{\bsnm{Moulton}, \binits{F.R.}}:
\bbtitle{An Introduction to Celestial Mechanics}.
\bpublisher{Dover Publications},
\blocation{New York}
(\byear{1914})
\end{bbook}
\endbibitem

\bibitem[\protect\citeauthoryear{Murray and Dermot}{1999}]{Murray&Dermot}
\begin{bbook}
\bauthor{\bsnm{Murray}, \binits{C.D.}},
\bauthor{\bsnm{Dermot}, \binits{S.F.}}:
\bbtitle{Solar System Dynamics}.
\bpublisher{Cambridge University Press},
\blocation{Cambridge}
(\byear{1999})
\end{bbook}
\endbibitem

\bibitem[\protect\citeauthoryear{Pathak et~al.}{2016}]{Niraj}
\begin{barticle}
\bauthor{\bsnm{Pathak}, \binits{N.}},
\bauthor{\bsnm{Sharma}, \binits{R.K.}},
\bauthor{\bsnm{Thomas}, \binits{V.O.}}:
\bjtitle{International Journal of Astronomy and Astrophysics}
\bvolume{6},
\bfpage{175}
(\byear{2016}).
doi:\doiurl{10.4236/ijaa.2016.62015}
\end{barticle}
\endbibitem

\bibitem[\protect\citeauthoryear{Plummer}{1919}]{Plummer}
\begin{barticle}
\bauthor{\bsnm{Plummer}, \binits{H.C.}}:
\bjtitle{The American Mathematical Monthly}
\bvolume{26}(\bissue{6})
(\byear{1919}).
doi:\doiurl{10.2307/2973529}
\end{barticle}
\endbibitem

\bibitem[\protect\citeauthoryear{Pollard}{1966}]{Pollard}
\begin{bbook}
\bauthor{\bsnm{Pollard}, \binits{H.}}:
\bbtitle{Mathematical Introduction to Celestial Mechanics}.
\bpublisher{Prentice Hall},
\blocation{New Jersey}
(\byear{1966})
\end{bbook}
\endbibitem

\bibitem[\protect\citeauthoryear{Poynting}{1903}]{Poynting1903}
\begin{barticle}
\bauthor{\bsnm{Poynting}, \binits{J.H.}}:
\bjtitle{Monthly Notices of the Royal Astronomical Society}
\bvolume{64}(\bissue{1}),
\bfpage{1}
(\byear{1903}).
doi:\doiurl{10.1093/mnras/64.1.1a}
\end{barticle}
\endbibitem

\bibitem[\protect\citeauthoryear{Pushparaj and
  Sharma}{2016}]{Pushparaj&Sharma2016}
\begin{barticle}
\bauthor{\bsnm{Pushparaj}, \binits{N.}},
\bauthor{\bsnm{Sharma}, \binits{R.K.}}:
\bjtitle{International Journal of Astronomy and Astrophysics}
\bvolume{6},
\bfpage{347}
(\byear{2016}).
doi:\doiurl{10.4236/ijaa.2016.64029}
\end{barticle}
\endbibitem

\bibitem[\protect\citeauthoryear{Richardson}{1980}]{Richardson}
\begin{barticle}
\bauthor{\bsnm{Richardson}, \binits{D.L.}}:
\bjtitle{Celestial Mechanics}
\bvolume{22},
\bfpage{231}
(\byear{1980}).
doi:\doiurl{10.1007/BF01229511}
\end{barticle}
\endbibitem

\bibitem[\protect\citeauthoryear{Robertson and Russell}{1937}]{Robertson}
\begin{barticle}
\bauthor{\bsnm{Robertson}, \binits{H.P.}},
\bauthor{\bsnm{Russell}, \binits{H.N.}}:
\bjtitle{Monthly Notices of the Royal Astronomical Society}
\bvolume{97}(\bissue{6}),
\bfpage{423}
(\byear{1937}).
doi:\doiurl{10.1093/mnras/97.6.423}
\end{barticle}
\endbibitem

\bibitem[\protect\citeauthoryear{Roy}{2005}]{Roy}
\begin{bbook}
\bauthor{\bsnm{Roy}, \binits{A.E.}}:
\bbtitle{Orbital Motion}.
\bpublisher{Institute of Physics Publishing},
\blocation{UK}
(\byear{2005})
\end{bbook}
\endbibitem

\bibitem[\protect\citeauthoryear{Schuerman}{1980}]{Schuerman}
\begin{barticle}
\bauthor{\bsnm{Schuerman}, \binits{D.W.}}:
\bjtitle{The Astrophysical Journal}
\bvolume{238},
\bfpage{337}
(\byear{1980}).
doi:\doiurl{10.1086/157989}
\end{barticle}
\endbibitem

\bibitem[\protect\citeauthoryear{Sharma}{1987}]{Sharma1987}
\begin{barticle}
\bauthor{\bsnm{Sharma}, \binits{R.K.}}:
\bjtitle{Astrophysics and Space Science}
\bvolume{135}(\bissue{2}),
\bfpage{271}
(\byear{1987})
\end{barticle}
\endbibitem

\bibitem[\protect\citeauthoryear{Simmons et~al.}{1985}]{Simmons}
\begin{barticle}
\bauthor{\bsnm{Simmons}, \binits{J.F.L.}},
\bauthor{\bsnm{McDonald}, \binits{A.J.C.}},
\bauthor{\bsnm{Brown}, \binits{J.C.}}:
\bjtitle{Celestial Mechanics}
\bvolume{35}(\bissue{145}),
\bfpage{146}
(\byear{1985}).
doi:\doiurl{10.1007/BF01227667}
\end{barticle}
\endbibitem

\bibitem[\protect\citeauthoryear{Szebehely}{1967}]{Szebehely}
\begin{bbook}
\bauthor{\bsnm{Szebehely}, \binits{V.}}:
\bbtitle{Theory of Orbits. The Restricted Problem of Three Bodies}.
\bpublisher{Academic Press},
\blocation{New York}
(\byear{1967})
\end{bbook}
\endbibitem

\bibitem[\protect\citeauthoryear{Thurman and Worfolk}{1996}]{Thurman}
\begin{bbook}
\bauthor{\bsnm{Thurman}, \binits{R.}},
\bauthor{\bsnm{Worfolk}, \binits{P.A.}}:
\bbtitle{The Geometry of Halo Orbits in the Circular Restricted Three-body
  Problem}.
\bpublisher{Technical Report},
\blocation{Minneapolis}
(\byear{1996})
\end{bbook}
\endbibitem

\bibitem[\protect\citeauthoryear{Tiwary and Kushvah}{2015}]{Tiwary2015}
\begin{barticle}
\bauthor{\bsnm{Tiwary}, \binits{R.D.}},
\bauthor{\bsnm{Kushvah}, \binits{B.S.}}:
\bjtitle{Astrophysics and Space Science}
\bvolume{357}(\bissue{1}),
\bfpage{73}
(\byear{2015}).
doi:\doiurl{10.1007/s10509-015-2243-5}
\end{barticle}
\endbibitem

\bibitem[\protect\citeauthoryear{Vallado}{2013}]{Vallado}
\begin{bbook}
\bauthor{\bsnm{Vallado}, \binits{D.A.}}:
\bbtitle{Fundamentals of Astrodynamics and Applications}.
\bpublisher{Microcosm},
\blocation{Hawthorne}
(\byear{2013})
\end{bbook}
\endbibitem

\bibitem[\protect\citeauthoryear{Winter}{1941}]{Winter}
\begin{bbook}
\bauthor{\bsnm{Winter}, \binits{A.}}:
\bbtitle{The Analytical Foundations of Celestial Mechanics}.
\bpublisher{Princeton University Press},
\blocation{New Jersey}
(\byear{1941})
\end{bbook}
\endbibitem

\end{thebibliography}

\clearpage

%

\end{document}